\documentclass[aps,prb,twocolumn,showpacs,preprintnumbers,notitlepage,superscriptaddress,floats]{revtex4-2}
\usepackage{youngtab}
\usepackage{ifthen}
\usepackage{graphicx}
\usepackage{amsmath}
\usepackage{amssymb}
\usepackage{amsfonts}
\usepackage{braket,slashed}
\usepackage[colorlinks=true, breaklinks=true, linkcolor=red, citecolor=blue, urlcolor=blue]{hyperref} 
\usepackage{xcolor}

\newcommand{\overbar}[1]{\mkern 1.5mu\overline{\mkern-1.5mu#1\mkern-1.5mu}\mkern 1.5mu}

\newcommand{\su}[3]{
  \ifthenelse{\equal{#2}{0}}{\ifthenelse{\equal{#1}{0}}{}{\mult{#1}}\overset{\bf #3}{\underset{\phantom{.}}{\bullet}}}{\ifthenelse{\equal{#1}{0}}{}{\mult{#1}}\underset{\phantom{.}}{\overset{\bf #3}{\Yboxdim9pt\yng(#2)}}}
}

\newcommand{\mult}[1]{{ \fcolorbox{gray!70}{gray!70}{\textcolor{white}{\footnotesize \bf #1}}}\;}

\newcommand{\mydiagram}[1]{\quad\vcenter{\hbox{\includegraphics[scale=0.4,page=#1]{Figures/diagrams.pdf}}}\quad}

\allowdisplaybreaks[1]

\newcommand{\code}[1]{\texttt{#1}}

\newcommand{\e}{\ensuremath{\mathrm{e}}}

\newcommand{\SU}{\ensuremath{\mathrm{SU}}}

\newcommand{\Z}{\ensuremath{\mathbb{Z}}}

\begin{document}

\title{Phase diagram of the chiral SU(3) antiferromagnet on the kagome lattice}
\author{Yi Xu}
\altaffiliation{These authors contributed equally to this work.}
\affiliation{Department of Physics and Astronomy, Rice University, Houston, TX 77005, USA}

\author{Sylvain Capponi}
\altaffiliation{These authors contributed equally to this work.}
\affiliation{Laboratoire de Physique Th\'eorique, Universit\'e de Toulouse, CNRS, UPS, France}

\author{Ji-Yao Chen}
\altaffiliation{These authors contributed equally to this work.}
\affiliation{Guangdong Provincial Key Laboratory of Magnetoelectric Physics and Devices, Center for Neutron Science and Technology, School of Physics, Sun Yat-sen University, Guangzhou, 510275, China}

\author{Laurens Vanderstraeten}
\affiliation{Department of Physics and Astronomy, University of Ghent, Krijgslaan 281, 9000 Gent, Belgium}

\author{Juraj Hasik}
\affiliation{Institute for Theoretical Physics, University of Amsterdam,
Science Park 904, 1098 XH Amsterdam, The Netherlands}

\author{Andriy H. Nevidomskyy}
\affiliation{Department of Physics and Astronomy, Rice University, Houston, TX 77005, USA}

\author{Matthieu Mambrini}
\affiliation{Laboratoire de Physique Th\'eorique, Universit\'e de Toulouse, CNRS, UPS, France}

\author{Karlo Penc}
\affiliation{Institute for Solid State Physics and Optics, Wigner Research Centre for Physics, H-1525 Budapest, P.O.B. 49, Hungary}

\author{Didier Poilblanc}
\affiliation{Laboratoire de Physique Th\'eorique, Universit\'e de Toulouse, CNRS, UPS, France}

\begin{abstract}
Motivated by the search for chiral spin liquids (CSL), we consider a simple model defined on the kagome lattice of interacting SU(3) spins (in the fundamental representation) including two-site and three-site permutations between nearest neighbor sites and on triangles, respectively. By combining analytical developments and various numerical techniques, namely exact Lanczos diagonalizations and tensor network variational approaches, we find a rich phase diagram with non-topological (``trivial") and topological (possibly chiral) gapped spin liquids (SLs). Trivial spin liquids include an Affleck-Kennedy-Lieb-Tasaki (AKLT)-like phase and a trimerized phase, the latter breaking the inversion center between the up and down triangles of the kagome lattice. A topological SL is stabilized in a restricted part of the phase diagram by the time-reversal symmetry breaking (complex) 3-site permutation term.
Analyzing the chiral edge modes of this topological SL on long cylinders or on finite disks, we have come up with two competing scenarios, either a CSL or a double Chern-Simon SL characterized by a single or by two counter-propagating Wess-Zumino-Witten SU(3)$_1$ chiral mode(s), respectively. In the vicinity of the extended ferromagnetic region we have found a magnetic phase corresponding either to a modulated canted ferromagnet or to a uniform partially magnetized ferromagnet.
\end{abstract}

\maketitle

%%%%%%%%%%%%%%%%%%
\section{Introduction}
%%%%%%%%%%%%%%%%%%

The electronic and magnetic properties of materials in Nature arise from the interactions between SU(2)-symmetric fermions - the electrons, on the background of nuclei. 
When the interactions are strong, such materials can be well described by the electronic Hubbard model, which has been extensively studied in the field of condensed matter physics. Recent developments venture beyond this SU(2) paradigm. In one form, through emergent SU(4) symmetry conjectured for instance in strong spin-orbit materials~\cite{Yamada2018} or twisted bilayer graphene~\cite{Chichinadze2022}. Yet a more direct approach, facilitated by the  continuous progress in ultracold atom platforms, is to emulate the physics of the SU(N)-symmetric Hubbard model by loading N-color atoms onto optical lattices~\cite{Cazalilla2014,Gorshkov2010}. These experimental platforms provide an ideal environment for exploring and engineering exotic phases that have yet to be discovered in real materials.

Among exotic phases of matter, topological spin liquids (TSL) have been the subject of intense experimental and theoretical research activities since the seminal SU(2)-symmetric Resonating Valence Bond (RVB) proposal by Anderson and Fazekas~\cite{Anderson1973,Fazekas1974}. Later on, the topological nature~\cite{Wen1990} of the RVB state on non-bipartite lattices has been noticed, and turned out to be transparent within the tensor network framework~\cite{Poilblanc2012}. 
However, parent Hamiltonians for such states, although local, are not physical involving complicated multi-site interactions~\cite{Schuch2012}. Hence it is not clear whether (non-chiral) TSL can be hosted in simple physical models. 
On the experimental side, platforms of Rydberg atoms~\cite{Browaeys2020,Semeghini2021,Giudici2022} offer beautiful implementations e.g. of the RVB physics of (hardcore) dimers. The realisation of synthetic gauge fields in cold atom platforms sets the stage to experimental observations of fractional Chern insulators or chiral spin liquids (CSL)~\cite{Chen2016,Weber2022,Leonard2022}. 

CSL define a broad class of TSL that break (spontaneously or not) time reversal and reflection (R) symmetries while preserving their product. Simple constructions of CSLs by Wen~\cite{WXG1989} and Kalmeyer and Laughlin~\cite{Kalmeyer1987} established a more precise connection to the physics of the fractional quantum Hall (FQH) effect. Evidence for spin-1/2 CSL has been provided in (frustrated) Heisenberg models in the presence of an additional chiral 3-site interaction, e.g. on the kagome lattice~\cite{Bauer2014,Gong2014a}, on the triangular lattice~\cite{Gong2017,Wietek2017,Cookmeyer2021} or on the square lattice~\cite{Nielsen2013,Poilblanc2017b}, and also in Hubbard insulators~\cite{SA2020,Boos2020}.

Preliminary investigations by one of the authors (K.P.)~\cite{Penc2022} have shown that a simple time reversal symmetric Hamiltonian on the kagome lattice consisting of (real) permutations of SU(3) spins (in the fundamental $\bf 3$ irreducible representation) on the triangular units admits an Affleck-Kennedy-Lieb-Tasaki (AKLT)-like~\cite{Affleck1987} state as the exact ground state. Interestingly, such a state bears a particularly simple tensor network representation (typical of AKLT states) involving $\bf{\overline 3}$ virtual particles fusing into singlets on every triangle. Also, the gapped AKLT phase has been shown to be stable under the addition of a nearest-neighbor Heisenberg coupling (two-site permutation of either positive or negative amplitude), limited by two critical points defined by equal amplitudes of the two and three site permutations~\cite{Penc2022}. Motivated by the recent quest for TSL and, in particular, for novel CSL we have extended KP's model by including time reversal symmetry-breaking (pure imaginary) triangular permutations providing a two-dimensional parameter manifold. A thorough investigation of the phase diagram of this model has been undertaken. Note that the same model has been studied using parton wavefunctions in a restricted domain of parameter space~\cite{WuTu2016} claiming the existence of an Abelian CSL. 

The paper is organized as follow; first, the model and the numerical methods are  described in Sec.~\ref{sec:model}. Then, the overall phase diagram is depicted in Sec.~\ref{sec:phase_diag} with a description of the various phases coming into play. In a second step, the ground states  and low-energy excitations of interesting phases of the phase diagram -- obtained by complementary numerical techniques -- are analysed in Sec.~\ref{sec:low_energy}. Interestingly, the model is shown to host a topological spin liquid phase in an extended domain of parameters. To characterize the nature of this TSL we have investigated the physics of the edge modes by different means, in particular using an accurate tensor network ansatz of the ground state. Details on analytical methods and numerical techniques such as Lanczos exact diagonalisations (ED), Matrix Product States (MPS) on cylinders and tensor network techniques using Projected Entangled Pair States (PEPS) and Projected Entangled Simplex States (PESS) 
 are provided in Appendix~\ref{appendix:analytic}, Appendix~\ref{appendix:ED}, Appendix~\ref{appendix:MPS} and Appendix~\ref{appendix:PESS}, respectively.

%%%%%%%%%%%%%%%%%%
\section{Model and numerical tools}\label{sec:model}
%%%%%%%%%%%%%%%%%%

The local Hilbert space on each site $i$ of the two-dimensional kagome lattice consists of the three states $|\alpha\rangle_i=\{A,B,C\}$ representing the fundamental (defining) representation $\bf 3$ of SU(3) (see Appendix~\ref{appendix:analytic-su3} for details on the SU(3) group). The interaction between these SU(3) ``spins'' is described by the SU(3)-symmetric Hamiltonian as follows:
\begin{eqnarray}
	\label{eq:model}
	H &=& J\sum_{\langle i,j\rangle}P_{ij}  + K_{R}\sum_{\triangle ijk}(P_{ijk} + P_{ijk}^{\,\,\,\,\,-1}) 
	\\ & + & iK_{I}\sum_{\triangle ijk }(P_{ijk} - P_{ijk}^{\,\,\,\,\,-1}),\nonumber
\end{eqnarray}
where the first term corresponds to two-site permutations over all nearest-neighbor bonds, and the second and third terms are the three-site permutations on all triangles, clockwise ($P_{ijk}$) and counterclockwise ($P_{ijk}^{-1}$). Written explicitly, $P_{ij}$ and $P_{ijk}$ are defined through their action on the local basis states, $P_{ij}|\alpha\rangle_i |\beta\rangle_j = |\beta\rangle_i |\alpha\rangle_j $  and $P_{ijk}|\alpha\rangle_i |\beta\rangle_j |\gamma\rangle_k= |\gamma\rangle_i |\alpha\rangle_j |\beta\rangle_k$, for a fixed orientation of the triangle $i$, $j$, $k$, say clockwise.  The $K_I$ term is ``chiral", in the sense that it breaks time reversal and reflection symmetries without breaking their product.

For convenience, in the following we use the parametrization on a sphere:
\begin{eqnarray}
	J&=&\cos \theta \cos \phi, \nonumber\\
	K_{R}&=&\cos \theta\sin\phi, \label{eq:paras}\\
	K_{I}&=&\sin\theta,\nonumber
\end{eqnarray}
where $0\le\phi<2\pi$ and it is sufficient to consider $\theta$ in the interval $[0,\pi/2]$  because of the symmetry $K_{I}\leftrightarrow -K_{I}$. Hence only the upper hemisphere of the parameter space is considered.

We have addressed the phase diagram of this model by complementary numerical tools. Lanczos ED on small periodic 21-site and 27-site clusters (torus geometry) have been performed to obtain the low-energy spectrum from which useful information on the nature of the phase can be extracted. Such clusters accommodate the SU(3) singlet subspace (and hence can describe spin liquids) and all available space group symmetries are used to label the many-body eigenstates. MPS calculations with explicit SU(3) symmetry on infinitely-long cylinders with finite circumference have also been performed to compute ground state properties \cite{ZaunerStauber2018}, entanglement spectra \cite{CL2013} and construct excitation ans\"atze \cite{VanDamme2021}. PEPS~\cite{Verstraete2004b} and PESS~\cite{Schuch2012,Xie2014} tensor networks have been considered and contracted on the infinite lattice using Corner Transfer Matrix Renormalization Group (CTMRG)~\cite{Nishino1996,Orus2012}. Both unconstrained and symmetric PEPS/PESS have been employed and variationally optimized e.g. using conjugate gradient optimization schemes~\cite{Vanderstraeten2016, Corboz2016}. While fully unconstrained or Abelian-symmetric (employing just $\mathrm{U}(1)\times \mathrm{U}(1)$ subgroup of SU(3) symmetry group) ans\"atze are less ``biased'' by construction, their optimization is more difficult and greatly benefits from an automatic differentiation procedure~\cite{Liao2017a,Hasik2021}.
In contrast, SU(3)-symmetric PEPS/PESS~\cite{Mambrini2016} depends on a much smaller number of variational parameters which can be optimized by numerically estimating the gradient vector~\cite{Poilblanc2017a,Poilblanc2017b}. Symmetric PEPS/PESS encoding SU(3)-rotation symmetry and translation invariance are particularly well-suited to describe singlet phases of matter~\cite{Kurecic2019,Chen2020}, where the SU(3) symmetry (implemented with the QSpace library~\cite{Weichselbaum2012,Weichselbaum2020}) also allows us to reach unusually large bond dimensions.

\begin{figure}[htb]
	\centering
    \includegraphics[width=\columnwidth]{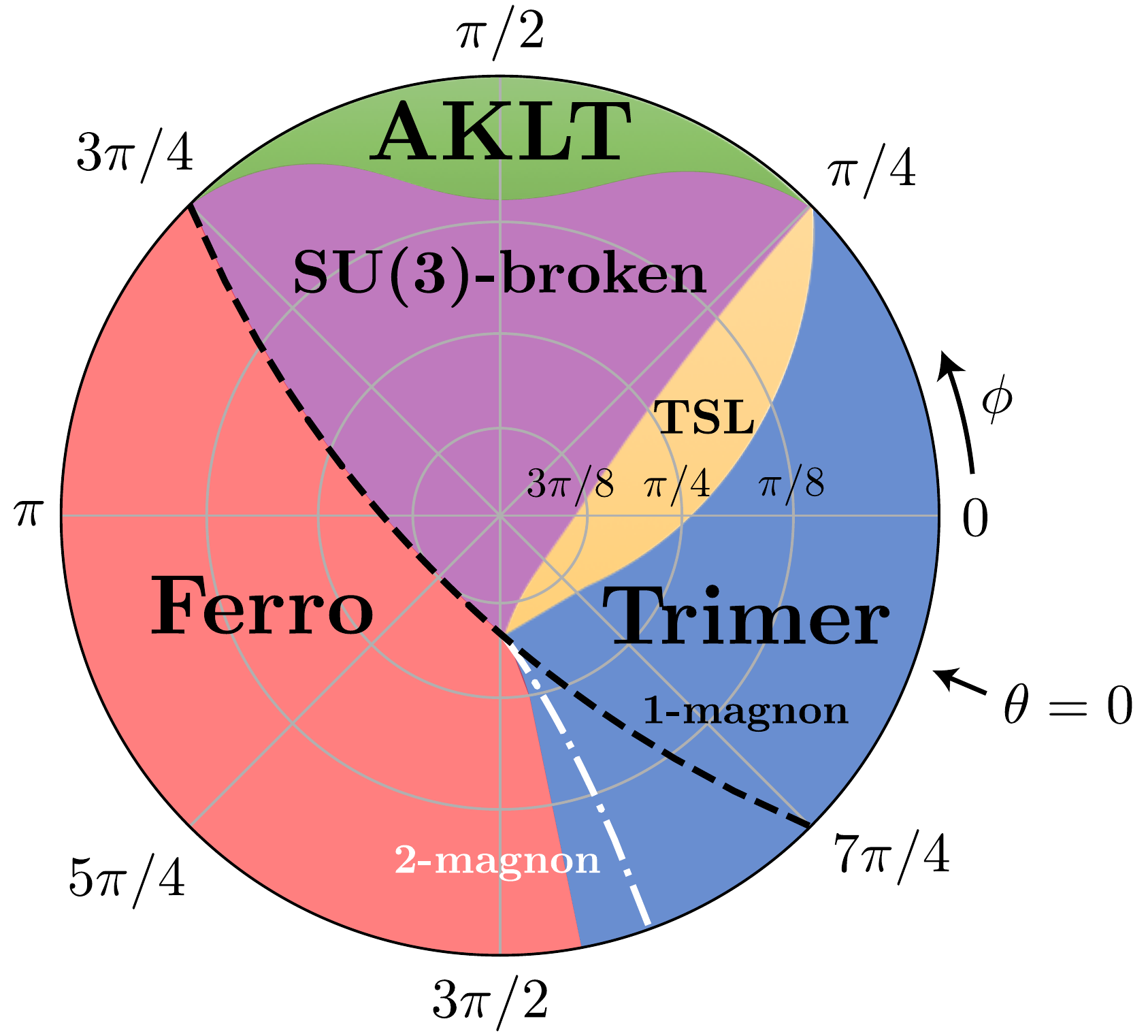}
	\caption{
   % \justifying
    Semi-quantitative phase diagram of the SU(3) chiral antiferromagnet on the kagome lattice using a stereographic projection (mapping the $(\theta,\phi)$ hemisphere onto a planar disc -- see Appendix~\ref{appendix:analytic-stereo}). 
		Shaded areas of different colors represent the various phases discussed in the text. The center of the plot ($\theta=\pi/2$) defines the ``North Pole" and the outer circle ($\theta=0$) the ``equator" parametrized by the azimuthal angle $\phi$, with the corresponding model parameters defined in Eq.~\ref{eq:paras}. The dashed (dash-dotted) line corresponds to the exact single-magnon (two-magnon) instability of the ferromagnetic phase. It is unclear whether the two-magnon instability gives the onset of the trimer phase. 
  }
		\label{fig:phasediag}
\end{figure}

%%%%%%%%%%%%%%%%%%
\section{Phase diagram} 
%%%%%%%%%%%%%%%%%%
\label{sec:phase_diag}
\subsection{Preliminaries}

The model studied here exhibits a very rich phase diagram as shown in Fig.~\ref{fig:phasediag}. The parameter space defined on a hemisphere is conveniently mapped onto a two-dimensional (2D) disk using a stereographic projection (see Appendix~\ref{appendix:analytic-stereo}). In this section we will describe qualitatively the phase diagram and the various phases we have encountered. More thorough descriptions, reports of the numerical results and discussions will be given in the subsequent sections.

To start with, it is useful to distinguish two types of phases: (i) the spin liquids (SL) whose ground states preserve both SU(3) rotational invariance (SU(3) singlets) and invariance under lattice translations -- like the Affleck-Kennedy-Lieb-Tasaki (AKLT) or chiral SL (CSL) phases -- but may break point group symmetry (like the trimer phase);  and (ii) the magnetically-ordered phases breaking the SU(3) rotational symmetry. The uniform fully-polarized ferromagnet is a trivial example of the latter type, but more complicated magnetic phases breaking lattice translation symmetry are also realised here. 
%Whether a third type of phases breaking translation invariance without breaking SU(3) is also realized in this model will be discussed later on. \anc{By this, do you mean some modulated trimer state? Or an AKLT state with a larger supercell?} 
Since the unit cell of the kagome lattice contains three sites and on each site there is a $\mathbf{3}$ spin, the Lieb-Schultz-Mattis (LSM) theorem~\cite{Lieb1961}, extended to higher spatial dimensions by Oshikawa~\cite{Oshikawa2000} and Hastings~\cite{Hastings2004}, and its generalization to SU(N)~\cite{Totsuka2017} does not apply to Hamiltonian (\ref{eq:model}) so that spin liquids in the phase diagram may or may not possess topological order. 
\begin{figure}[htb]
	\centering
		\includegraphics[width=\columnwidth]{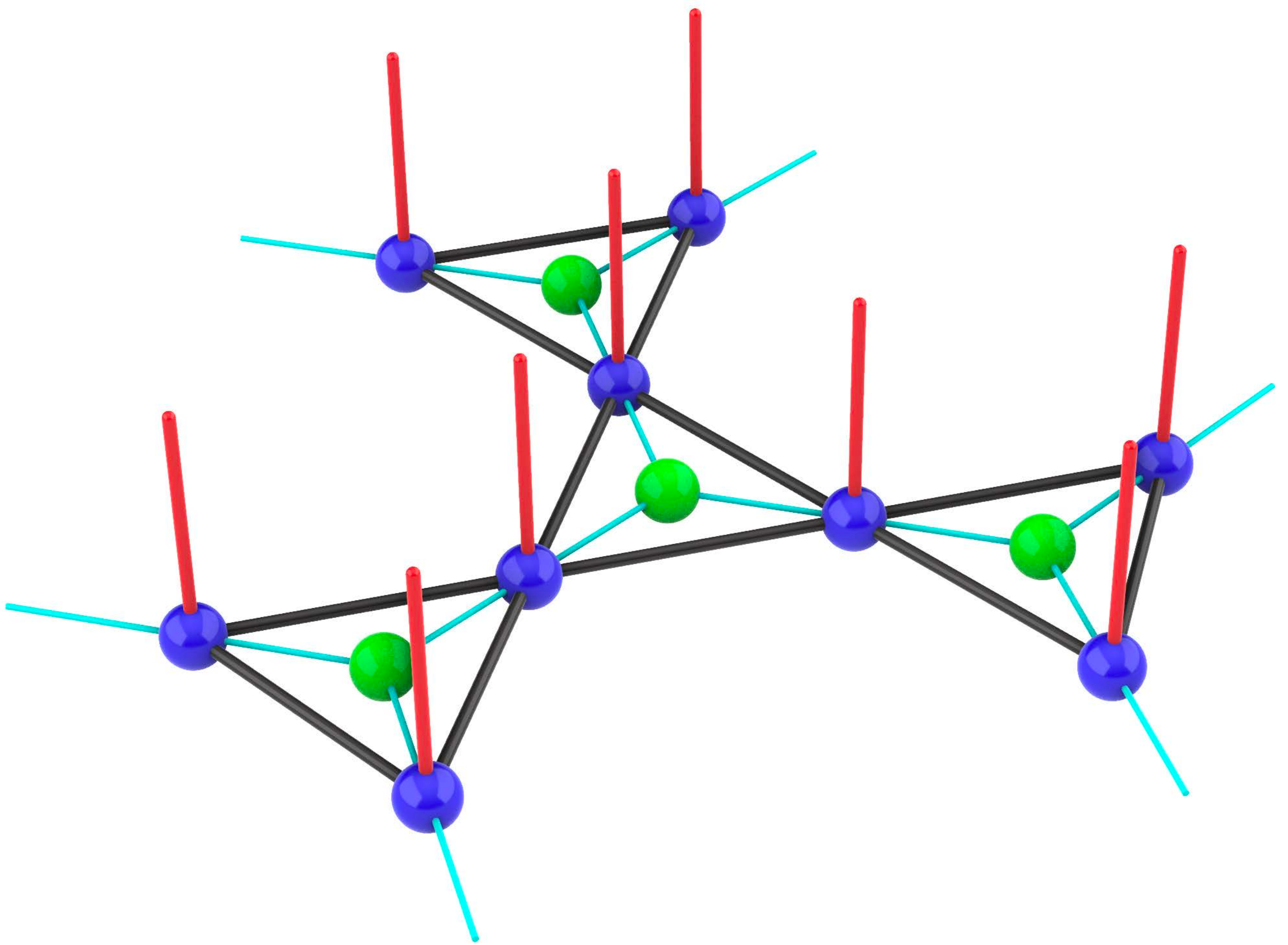}
	\caption{
    %\justifying
    PESS construction on the kagome lattice of corner-sharing triangles. Site-centered rank-3 tensors carrying the physical leg (in red) are represented in blue, while triangle-centered tensors represented in green fuse three virtual legs into an SU(3) singlet. In the case of SLs, the tensor network is uniform. PESS can also describe all other phases in the phase diagram with proper modifications to be discussed in the text and Appendix~\ref{appendix:PESS}. 
 \label{fig:AKLT-PESS}}
\end{figure}

Generically, the SL ground states can be conveniently defined/represented by a simple tensor network~\cite{Schuch2012,Poilblanc2012,Mambrini2016}. On the kagome lattice, the simplest version is a Projected Entangled Simplex State (PESS) (see Fig.~\ref{fig:AKLT-PESS}) involving a rank-3 tensor in each triangle (green sphere in Fig.~\ref{fig:AKLT-PESS}) and a rank-3 tensor with two virtual and one physical leg (blue sphere) on each site. Each bond connecting a site to the center of a triangle carries virtual SU(3) particles. The corresponding virtual Hilbert space $\cal V$ is therefore a direct sum of a certain number (labelled throughout by $D^*$) of SU(3) irreducible representations (irreps). On all triangles, the three virtual particles fuse into a singlet, and the trivalent tensor enforces the projection
${\cal V}^{\otimes 3}\rightarrow{\bf 1}$. 
On the sites, two virtual particles fuse to the physical state, and the site tensor enforces the projection
${\cal V}^{\otimes 2}\rightarrow{\bf 3}$. Here and throughout, we use the standard notation for the SU(3) irreps labeled by their dimension or, equivalently, by their Dynkin labels -- see Table~\ref{tab:tensors} in Appendix~\ref{appendix:MPS} for details. 
Besides the representation of spin liquids, the PESS formalism (associated to PEPS) turns out to be also extremely useful to investigate magnetic phases and phases breaking translation symmetry -- as will be discussed later on. Details about the PESS/PEPS constructions are given in Appendix~\ref{appendix:PESS}. 

\subsection{AKLT phase}

It has been shown by one of the authors (K.P.) that the non-chiral Hamiltonian defined by $K_I=0$ (i.e. $\theta=0$) in Eq.~\eqref{eq:model} has an exact ground state (GS) of the AKLT type in the range $\pi/4\le \phi\le 3\pi/4$~\cite{Penc2022}. It is closely related to the simplex solid of Arovas constructed in Ref.~\onlinecite{Arovas2008} which breaks no discrete lattice symmetry, but  we write the singlet creation operators using fermions, not bosons. This state can be represented by the simplest possible PESS representation just involving the unique irrep ($D^*=1$) ${\cal V}=\bar{\bf 3}$ on all virtual bonds. Hence, on all triangles three virtual $\overline{\bf 3}$ particles fuse into a singlet, $\overline{\bf 3}^{\otimes 3}\rightarrow{\bf 1}$ while, on the sites, two virtual $\bar{\bf 3}$ particles fuse into the physical irrep, $\overline{\bf 3}^{\otimes 2}\rightarrow{\bf 3}$. This construction provides a unique ground state and, since the AKLT phase is likely to be gapped (see later), we deduce that the AKLT phase is a featureless (or ``trivial") SL with no topological order. 

Being gapped, the AKLT phase survives when a sufficiently small chiral perturbation is introduced (i.e. $\theta<\theta_{\rm crit}(\phi)$, see later). To describe the AKLT ground state in these new regions of the phase diagram, one has to extend the previous PESS construction by increasing the (Hilbert space) dimension $D={\rm dim}({\cal V})$ of the virtual space. The singlet character of the GS implies that $\cal V$ is a direct sum of SU(3) irreps. The irreps appearing in this direct sum must fulfill a strict requirement to keep the same featureless (i.e., non-topological) character of the ground state. In fact, each irrep $\bf I$ of SU(3) is characterized by its $\mathbb{Z}_3$ charge $Q({\bf I})$ defined by the number of boxes of its Young tableau modulo 3 (e.g. $Q=2$ is equivalent to $Q=-1$). The AKLT PESS can only contain irreps $\bf I$ with the same charge as $\overline{\bf 3}$, i.e.  $Q({\bf I})=Q(\overline {\bf 3})=2$. Of course, the optimal choice of those irreps can only be determined numerically by a variational optimization scheme. Restricting to $D^*\leq 4$ irreps in the virtual space, we have found that 
${\cal V}=\overline{\bf 3}+\overline{\bf 3}+{\bf 6}+\overline{\bf 15}$ is the best choice.

\subsection{Trimer phase}

The SU(3) Heisenberg antiferromagnet on the kagome lattice (i.e. the $\phi=\theta=0$ point in our phase diagram) exhibits a trimer phase~\cite{Corboz2012b}, i.e. a simplex solid~\cite{Arovas2008} with different energy densities on the two types of up-pointing and down-pointing triangles (named \textit{up} and \textit{down} triangles hereafter). Hence, such a phase spontaneously breaks the (site) inversion center, resulting in a doubly-degenerate SU(3) singlet groundstate manifold.
We have shown that this phase survives in a rather extended region of our phase diagram. 

Similar to the AKLT phase, a particularly simple prescription exists for constructing PESS ans\"atze of the trimer phase by using different virtual spaces ${\cal V}_{\rm up}$ and ${\cal V}_{\rm down}$ for the up and down triangles, respectively. Let us start with the extreme case of decoupled SU(3) singlets on,  say, up triangles. An exact PEPS representation is given by 
${\cal V}_{\rm up}={\bf 3}$ and ${\cal V}_{\rm down}={\bf 1}$ and the corresponding (unique) $C_{\rm 3v}$-symmetric trivalent tensors on the up and down triangles encode the fusion rules 
${\bf 3}^{\otimes 3}\rightarrow {\bf 1}$ 
and ${\bf 1}^{\otimes 3}\rightarrow {\bf 1}$, respectively. Also, the site tensors encode the trivial fusion ${\bf 3}\otimes {\bf 1}\rightarrow {\bf 3}$. We note that the two irreps of the up and down virtual spaces have different ${\mathbb Z}_3$ charges, $Q_{\rm up}=1$ and $Q_{\rm down}=0$, respectively. This suggests that the PESS ansatz of a generic trimer state in which the up triangles are entangled can be constructed by simply adding more irreps of the same ${\mathbb Z}_3$ charge in ${\cal V}_{\rm up}$ and ${\cal V}_{\rm down}$. Restricting to $D^*=2$ irreps we found that ${\cal V}_{\rm up}={\bf 3}+\overline{\bf 6}$ and ${\cal V}_{\rm down}={\bf 1}+{\bf 8}$ provide the best ansatz. Note that, in such a construction, the inversion center is obviously broken and a pair of ground states is obtained by simply switching the virtual spaces between the up and down triangles.

\subsection{Topological spin liquid}

We have also found a gapped topological spin liquid (TSL) stabilized in a significant region of the phase diagram provided the chiral $K_I$ term is present ($\theta\ne 0$) in Eq.~\eqref{eq:model}, as shown in Fig.~\ref{fig:phasediag}. 
The region of stability of this phase includes the parameters $K_R/J\simeq 0.6$, $K_I/J\simeq 0.45$ (i.e $\theta\sim 0.13\pi$ and $\phi\sim 0.17\pi$) proposed in \cite{WuTu2016} as the optimal parameters for the stability of an Abelian CSL. 
Such a phase does not break any lattice symmetry (it is perfectly uniform), nor does it break the SU(3) symmetry. Moreover, it possesses topological order as defined by Wen~\cite{Wen1990,wen-review2017}. 
Interestingly, the existence of topological order is not {\it a priori} guaranteed in SU(3) kagome spin liquids since the LSM theorem does not apply here, as already noted above. 
Then, the ground state degeneracy is expected to be associated to the 3 possible values of the $\mathbb{Z}_3$ charge. Indeed, ED (MPS) on a torus (on an infinite cylinder) reveals 3 quasi-degenerate ground states in some extended region of the phase diagram.

A faithful description of such a phase in terms of PESS is in fact possible. A necessary (but not always sufficient) condition for the existence of (at least) 3 topological sectors on the infinite cylinder is that the virtual space should contain at least one irrep within each of the 3 $\mathbb{Z}_3$ charge sectors. A minimal choice would then be ${\cal V}={\bf 1}+{\bf 3}+\overline{\bf 3}$. Below we show that increasing the virtual space to ${\cal V}={\bf 1}+{\bf 3}+\overline{\bf 3}+{\bf 6}+{\bf 8}$, with 
%two extra 
additional irreps of charge $Q=2$ and $Q=0$, provides an optimal low-energy ansatz of the TSL. 

As reported below in Section~\ref{sec:low_energy}, we find that this TSL phase exhibits chiral edge modes, as revealed by its entanglement spectrum (ES). The content of the edge modes is described in terms of a SU(3)$_1$ Wess-Zumino-Witten (WZW) Conformal Field Theory (CFT), and should fully characterize the nature of this Abelian TSL. The results obtained with our PESS ansatz show edge modes of both right- and left-moving chiralities (and different velocities) consistent with a SU(3)$_1$ doubled Chern-Simons (DCS) Topological Field Theory (TFT)~\cite{Kurecic2019, Arildsen2022b}. On the other hand, the ED and MPS results rather point towards a chiral spin liquid (CSL) phase exhibiting a single {\it chiral} edge mode. Later in Section~\ref{sec:low_energy} we shall discuss further the pros and cons for the CSL  or for the DCS phase.

\subsection{Ferromagnetic phase}
\label{sec:ferro_main}
The ferromagnet is 
%a fully symmetric 
a lattice-symmetric
state which spontaneously breaks the internal SU(3) symmetry,
where both the two-site and the three-site permutations act trivially with eigenvalues $+1$.  Hence the ground state energy per site is simply $e_F = 2 J + 4 K_R/3$.

To determine the phase boundary of the ferromagnetic state, we calculate the dispersion of a single magnon. By this, we mean a configuration such that the flavors are $A$ on all sites except one, which is, say, a $B$. The Hamiltonian then hops the $B$ flavor to neighboring sites, giving it a dispersion determined by the eigenvalues of the three-by-three matrix~(\ref{eq:magnon}) in Appendix~\ref{appendix:analytic-magnon}.  The matrix~(\ref{eq:magnon}) describes three magnon branches. If the dispersion of the magnon, measured from the energy of the ferromagnetic state, is negative, the ferromagnetic phase becomes unstable and ceases to exist.
Scrutinizing the dispersions, it turns out that they are functions of $J+K_R$ and $K_I^2$ only. The maximum and minimum of the dispersions are always at momentum $\mathbf{q}=0$, where the energies are $0$ and
$-6 (J + K_R) \pm 2 \sqrt{3} K_I$. We get a positive dispersion for 
\begin{equation}
   J + K_R <  - |K_I| /\sqrt{3}.
\end{equation}
Conversely, the ferromagnetic state is unstable for $J + K_R > - |K_I| /\sqrt{3}$. On the boundary, when $ J + K_R = - |K_I| /\sqrt{3}$, the $0$ energy band becomes flat. Localized modes on hexagons common to kagome lattice appear, but with amplitudes $e^{i j \pi/3}$  as we go around the hexagon -- the complex amplitudes reflect the chiral nature of the model. 

The one-magnon instability line is shown as a dashed line in Fig.~\ref{fig:phasediag}. Interestingly, we find numerically that, within the one-magnon stability region, the ferromagnetic state is nevertheless unstable with respect to the trimer phase.

 We have also envisioned the possibility of a two-magnon instability by considering a two-magnon excitation of the ferromagnetic state, where two spins are flipped with different flavors. Details of the calculations are given in section~\ref{appendix:analytic-2magnons} of Appendix~\ref{appendix:analytic}.
The two-magnon calculation aims to reveal whether the interaction between the magnons could be attractive and lead to a bound state. In that case, the boundary of the ferromagnetic phase would shrink further. This is indeed what has been found as shown in the phase diagram of Fig.~\ref{fig:phasediag}. Numerically, we found that this two-magnon instability line marks (approximately) the instability to the trimer phase. 

\begin{figure}[thb]
    \centering
    \includegraphics[width=0.9\columnwidth]{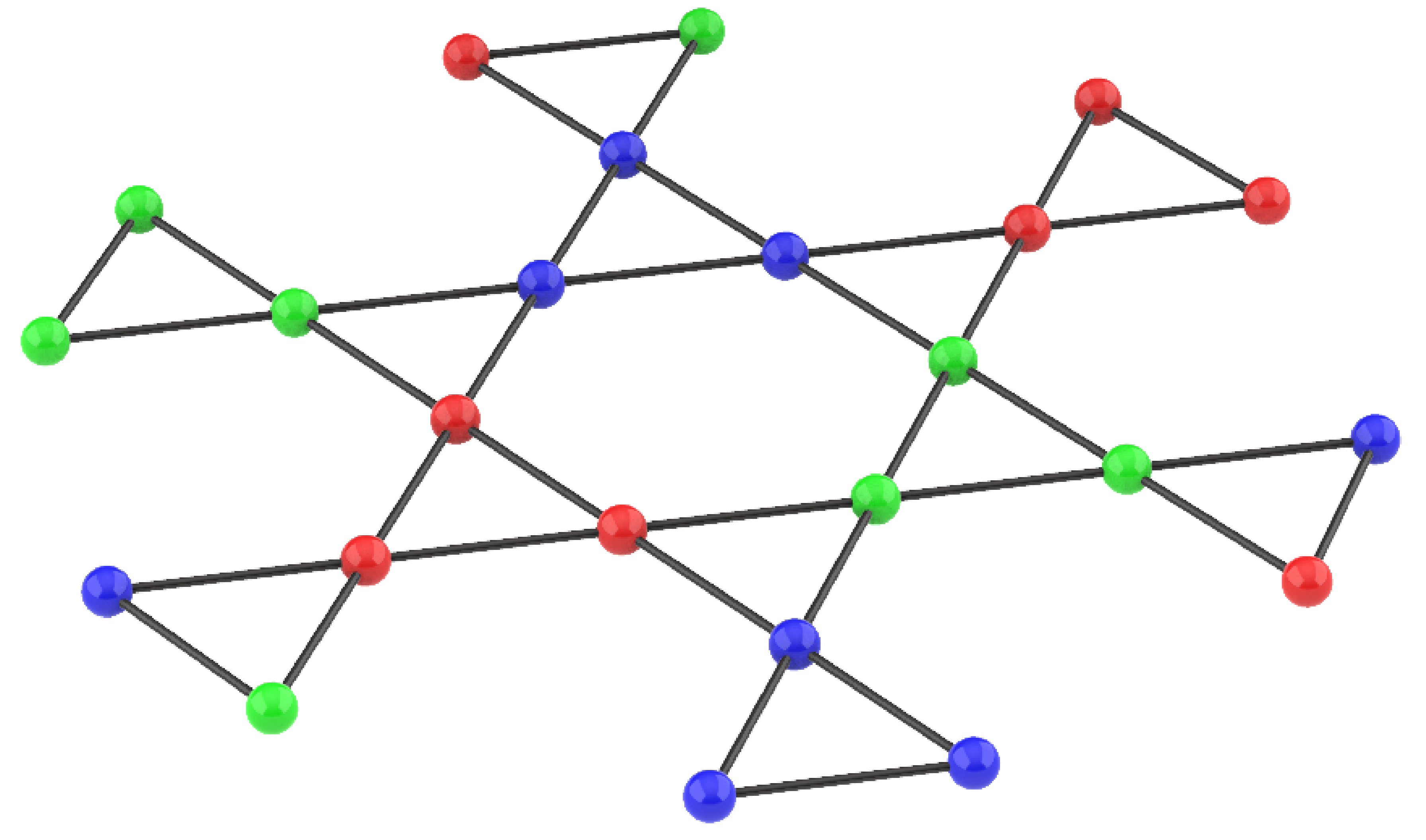}
    \caption{ 
       %\justifying
    9-site $\sqrt{3}\times \sqrt{3}$ unit cells tiled in C3-rotation symmetric patterns. The colors indicate which one of the three SU(3) colors has the dominant weight. Note that the colors in each, say, up triangle are identical and have the same dominant weight magnitude. }
    \label{fig:canted_ferro} 
\end{figure}

\subsection{SU(3)-broken phase}

When crossing the one-magnon instability line in the region roughly $\pi/4<\phi<3\pi/4$ (magenta color in Fig.~\ref{fig:phasediag}), the ferromagnetic state becomes unstable and gives rise to a partially magnetized phase (with magnetization $0<m<1$), hence breaking SU(3) symmetry. Such spontaneous SU(3)-breaking may occur while preserving or (spontaneously) breaking translation symmetry. 

The translation symmetry-broken phase is characterized by a spin canting occurring on three triangular sub-lattices separately, which requires a 9-site unit cell. The canted spins (all of the same length) on each sub-lattice form either a stripy pattern or a C3-rotation symmetric pattern, with all three sub-lattices having the same overall spin direction (see Appendix \ref{appendix:PESS}). In our calculations, the so-called C3-2 $\sqrt{3}\times\sqrt{3}$ C3-rotation symmetric pattern in which the site SU(3)-color occupations are identical in each (let say) up triangle - see Fig.~\ref{fig:canted_ferro} - seem to be energetically favored over the other C3-rotation symmetric or stripe patterns discussed in section \ref{app:3triangles} of Appendix~\ref{appendix:PESS}. 

The second competing magnetic phase can be described by a uniform translationally-invariant (3-site) PEPS, as discussed in section~\ref{app:1triangle} of Appendix~\ref{appendix:PESS}. After energy optimization, the magnetization in such {\it a priori} un-restricted ansatz turns out to be uniform and significantly lower than in the modulated C3-2 phase. Note also, the magnetizations on the three sites within the unit cell are not canted but rather collinear. Interestingly, the numerics point towards a jump of the magnetization at the boundary to the fully polarized phase. This is indeed expected from the analytic calculation of the one-magnon instability in section~\ref{appendix:analytic-magnon} of Appendix~\ref{appendix:analytic} predicting infinite compressibility at the transition.

\begin{figure}[!ht]
\begin{centering}
\includegraphics[width=0.97\columnwidth]{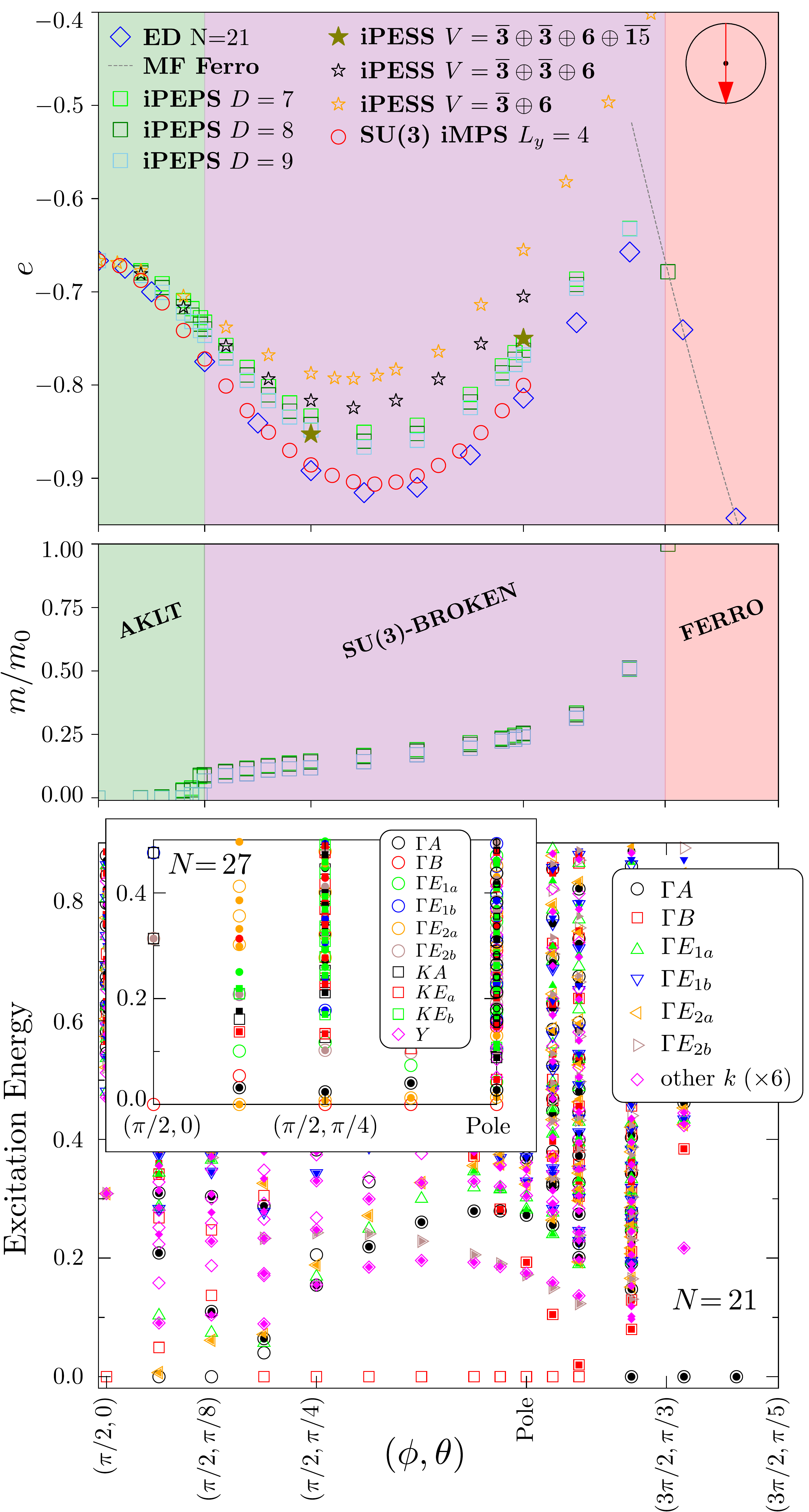}
\end{centering}
    \caption{
    %\justifying
    Top: Energetics of ED, iMPS, iPESS and iPEPS wave functions along the $\phi=\pi/2$ [mod $\pi$] meridian where $J=0$, i.e., along the vertical diameter of Fig.~\ref{fig:phasediag} as highlighted in the top right corner. The dashed line corresponds to the exact ferromagnet energy. The phase boundaries are approximate except for the canted ferro-ferro transition at $(\phi,\theta)=(3\pi/2,\pi/3)$. 
    Middle: Uniform magnetization of the unit cell $m$ 
    in units of $m_0$. 
    Bottom: ED low-energy (excitation) spectrum of a periodic $21$-site cluster. Open (closed) symbols show the singlet (non-singlet) eigenstates and the GS energy has been subtracted. Different symbols correspond to different momenta as shown in the legend. The black circles on the right correspond to the largest SU(3) irrep. Inset panel: ED on a 27-site torus shows the disappearance of the gapped region close to the pole.
		\label{fig:cut05phi}
  }
\end{figure}

\section{Ground states and low-energy excitations}
\label{sec:low_energy}

A crude determination of the phase diagram in Fig.~\ref{fig:phasediag} and of the boundaries of the various phases was first obtained by inspecting the low-energy ED spectra on a periodic 21-site torus (see Appendix~\ref{appendix:ED} for details). These results were then extended to a 27-site torus (for a much smaller set of parameters) and compared against  %confronted to
the results obtained by tensor network methods (MPS, iPESS, iPEPS) to refine the phase diagram. For simplicity, we shall here focus on three different cuts -- the $\phi=0$ [mod $\pi$] and the $\phi=\pi/2$ [mod $\pi$] meridians, together with a portion of the $\theta=\pi/8$ latitude -- which contain all the phases we have encountered. 

\subsection{Energetics}

\begin{figure}[!ht]
	\begin{centering}
 \includegraphics[width=\columnwidth]{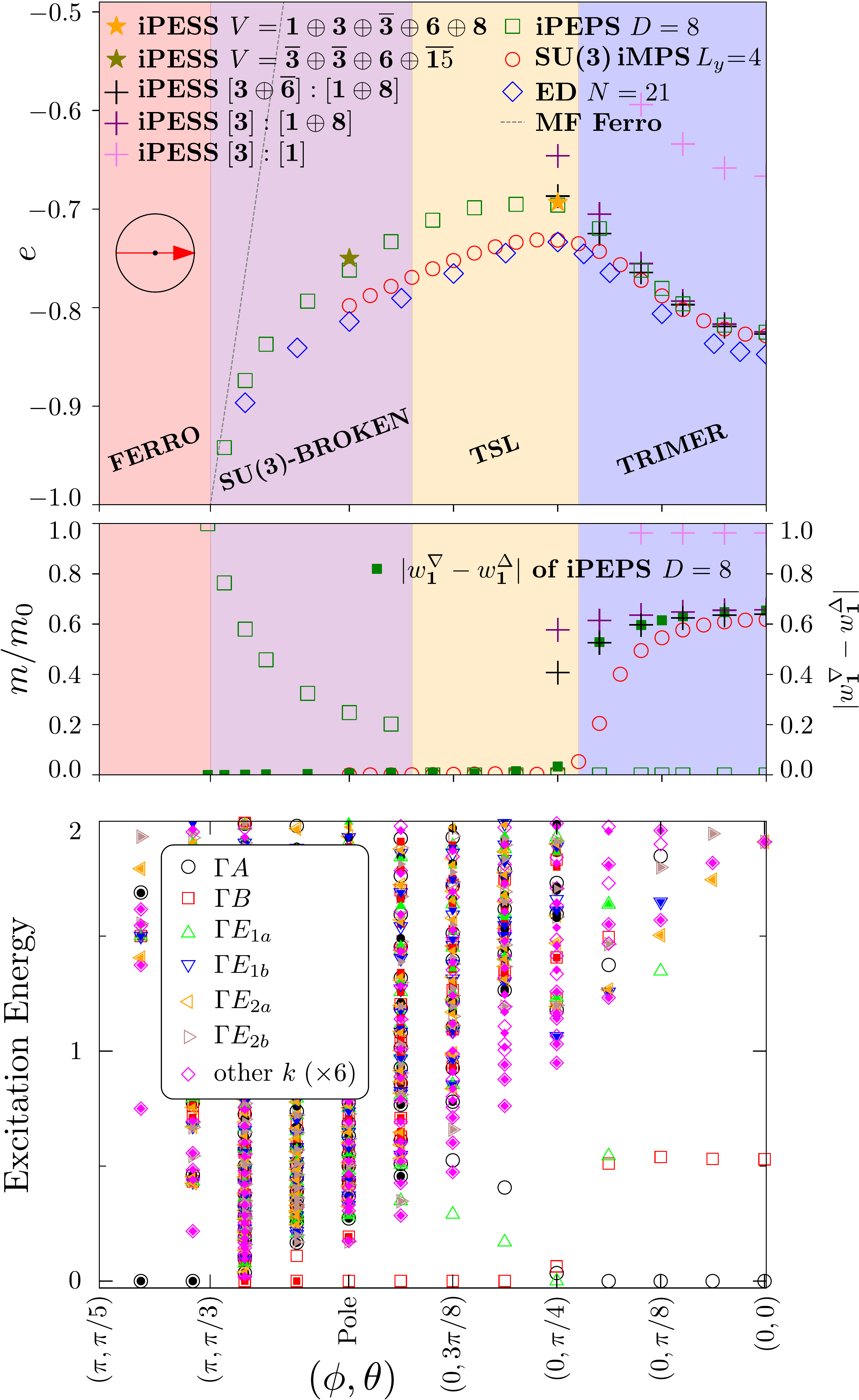}
\end{centering}
	\caption{
 %\justifying
 Top: Energetics of ED, iMPS, iPESS and iPEPS wave functions on the $\phi=0$  [mod $\pi$] meridian where $K_R=0$, i.e. from the leftmost point on the equator to the rightmost point on the equator via the North Pole in Fig.~\ref{fig:phasediag} as highlighted in the top right corner. The iPESS ansatz for the trimer phase is indicated in the legend as $[{\cal V}_{\rm up}]:[{\cal V}_{\rm down}]$.
Middle: Order parameters of iPEPS wave functions.  The uniform magnetization $m$ (open green squares) and its non-zero value identifies the SU(3)-broken phase. The trimer phase order parameter indicated by the arrow is shown on the right scale for various ansatze. 
 Bottom: ED low-energy (excitation) spectrum of a periodic $21$-site cluster.  The same symbols are used as in Fig.~\ref{fig:cut05phi}. 
		\label{fig:cut00phi}
	}
\end{figure}

\begin{figure}[!ht]
    \begin{centering}
     \includegraphics[width=0.98\columnwidth]{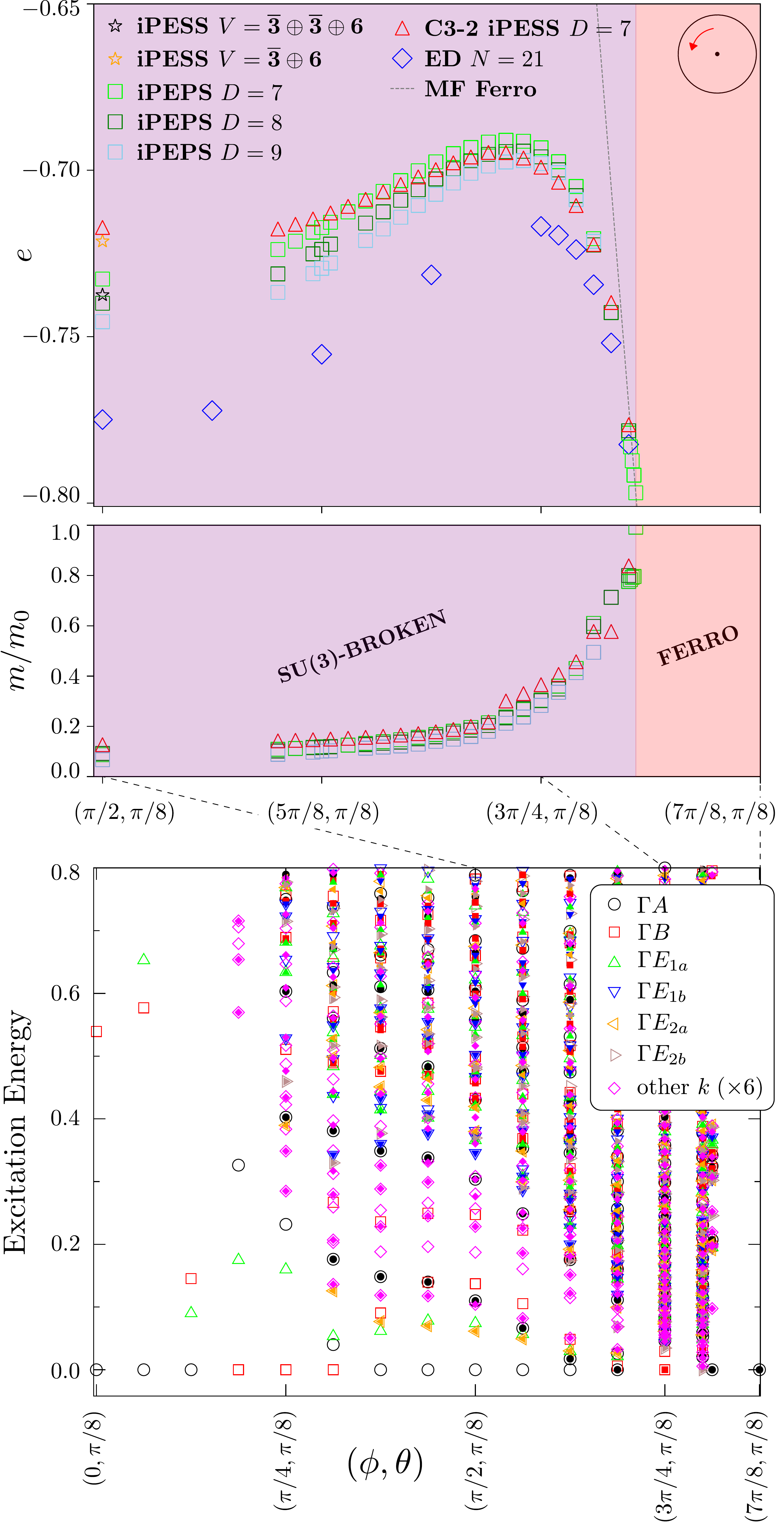}
	\end{centering}
	\caption{
  %\justifying
 Top: Energetics of ED, iPESS and iPEPS wave functions along part of the $\theta=\pi/8$  latitude as highlighted in the top right corner.
 Bottom: ED low-energy (excitation) spectrum of a periodic $21$-site cluster on the same latitude, but on a larger arc from $\phi=0$ (in the trimer phase).
 \label{fig:cut0125theta}
	}
\end{figure}

The top panels in  Figs.~\ref{fig:cut05phi}, \ref{fig:cut00phi} and \ref{fig:cut0125theta} show comparisons of the energy per site obtained by ED, iMPS, iPESS and iPEPS, along the aforementioned vertical, horizontal and circular cuts, respectively. The ED ground state energies have been all obtained from the same periodic 21-site cluster preserving all the symmetries of the infinite lattice. 
In Figs.~\ref{fig:cut05phi} and \ref{fig:cut00phi} the iMPS energy has been obtained on a finite width ($L_y=4$) cylinder and SU(3) symmetry.

We believe the ED and iMPS energies provide a (non-rigorous) lower bound of the energy due to strong finite-size effects. 

We have used translationally invariant SU(3)-symmetric iPESS calculations to target SU(3) singlet phases, like the AKLT phase (virtual spaces ${\cal V=}\bf\overline 3$, $\bf {\overline 3}+6$, $\bf {\overline 3}+{\overline 3}+ 6$, $\bf {\overline 3}+{\overline 3}+6+{\overline{15}}$) shown in Figs.~\ref{fig:cut05phi} and \ref{fig:cut00phi} and the TSL phase (${\cal V}=\bf {1}+{3}+{\overline 3}+6+{8}$) shown in Fig.~\ref{fig:cut00phi}. To describe the  trimer phase (see Fig.~\ref{fig:cut00phi}) one has to extend symmetric iPESS by allowing two different virtual spaces ${\cal V}_{\rm up}$ and ${\cal V}_{\rm down}$ on the up and down triangles, characterized by different $Z_3$ charges $Z_{\rm up} = 1$ and $Z_{\rm down} = 0$, respectively. In the region of stability of the trimer phase, we indeed observe that the $[{\cal V}_{\rm up}={\bf 3}+\overline{\bf 6}]:[{\cal V}_{\rm down}={\bf 1}+{\bf 8}]$ ansatz provides a very good variational energy, comparable to e.g. that of a generic $D=8$ iPEPS ansatz (whose properties will be discussed later on). 

In Fig.~\ref{fig:cut05phi}, moving away from the AKLT phase towards the (fully-polarized) ferromagnetic phase, we see that the optimization of unconstrained (1-triangle) iPEPS provides comparable, or even better, energies than the SU(3)-symmetric ansatze, although with a modest bond dimension ($D=7,8,9$ compared to e.g. $D=27$ for one of the SU(3)-symmetric ansatz), suggesting the existence of an intermediate phase breaking SU(3)-symmetry. This also happens in a limited region of Fig.~\ref{fig:cut00phi}. 
In Fig.~\ref{fig:cut0125theta}, in the vicinity of the (fully-polarized) ferromagnetic phase, the optimization of an unconstrained 
$D=7$ C3-2 iPESS provides good variational energies comparable to or even better than the previously mentioned 1-triangle iPEPS. This suggests that in the magnetic SU(3)-broken phase, the lattice translation symmetry may be spontaneously broken to form a 3-triangle $\sqrt{3}\times\sqrt{3}$ unit cell order, corresponding to the modulation specifically encoded in the C3-2 PESS ansatz.

\begin{figure}[htb]
	\centering
	\includegraphics[width=0.98\columnwidth]{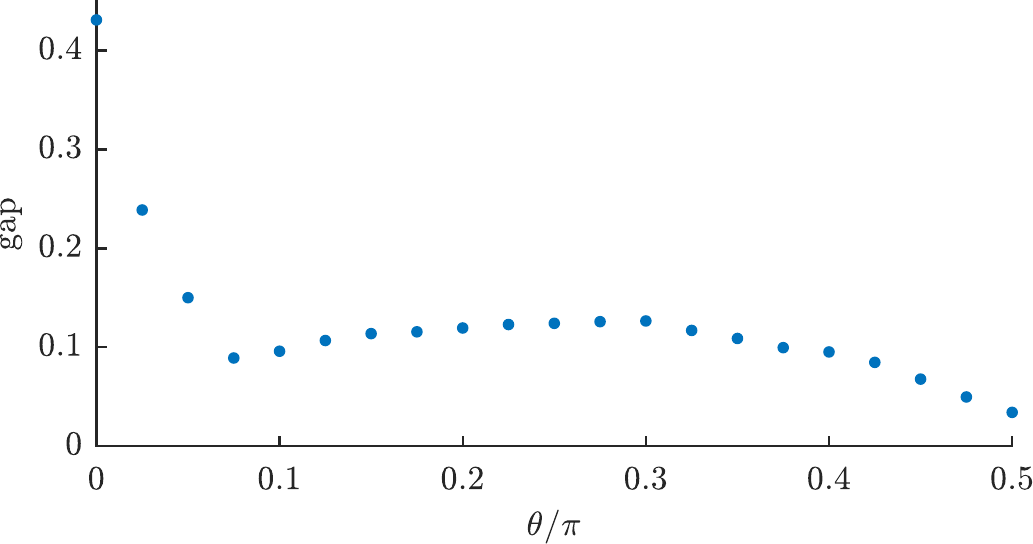}
	\caption{%\justifying AKLT gap along the $\phi=\pi/2$ [mod $\pi$] meridian ($J=0$), estimated by the MPS excitation ansatz on an $L_y=4$ cylinder. Extrapolating the rapid decrease of the gap on the left side gives a crude estimate of the critical value $\theta_c\simeq 0.1\pi$ of the transition between the AKLT and the (gapless) magnetic phases. The spurious gap found for $\theta >\theta_c$ is due to the constrained form of the MPS excitation ansatz. 
 \label{fig:aklt_gap_imps}}
\end{figure}

\subsection{Order parameters}

To further characterize the magnetic SU(3)-broken phase, we define the uniform magnetization of the unit cell $m = |\sum_{i\in\textrm{unit cell}} \vec{\lambda}_i|$, $\lambda_i^\alpha$ being the 8 generators of SU(3) acting on the site $i$, giving the fraction $m/m_0$ of the magnetization $m_0$ of the fully polarized ferromagnet. As shown in the middle panels of Figs.~\ref{fig:cut05phi}, \ref{fig:cut00phi} and \ref{fig:cut0125theta} the SU(3)-broken phase can be determined more accurately from the onsets of finite values of $m$ which reaches its maximal value $m_0=2/\sqrt{3}$ in the fully polarized ferromagnet. 

We have also defined the average magnitude as $\tilde{m} = \sum_{i\in\textrm{unit cell}} |\vec{\lambda}_i|$ (not shown). 
We observe that $\tilde{m}$ is different from $m$ only for the 3-triangle iPEPS signaling canting of the site polarizations. In contrast, the 1-triangle iPEPS shows aligned magnetizations on the 3 sites of the unit cell. Interestingly, the D=7 (1-triangle) iPEPS data point to a jump in the magnetization at the boundary
to fully polarized ferromagnet.

The product of three physical $\bf 3$-irreps can be decomposed into a direct sum of four irreps, given by the fusion rule ${\bf 3}^{\otimes 3}=\bf 1 + \bf 8 + \bf 8 + \bf 10$. Hence, for the three SU(3) spins on each triangle, one can define the projection operators onto corresponding irreps in the direct sum (weights of irreps), $w_{\bf 1, 8, 8, 10}$, which satisfy the completeness relation $w_{\bf 1}+w_{\bf 8}+w_{\bf 8}+w_{\bf 10}=1$. As the trimer states spontaneously break the inversion symmetry and form SU(3) trimers on either up or down triangles, we define the trimer order parameter as the difference between projections $w_{\bf 1}^{\nabla,\Delta}$ within the down and up triangles onto $\bf 1$-irrep (weight of $\bf 1$-irrep) to identify the trimer phase.
This trimer order parameter is shown on the middle panel of Fig.~\ref{fig:cut00phi} (right scale). Interestingly, the unconstrained $D=8$ iPEPS calculation gives a very similar order parameter in the trimer phase to the SU(3)-symmetric PESS ansatze specially designed to describe the trimer phase. It also shows an abrupt jump at the onset of the TSL phase while the iMPS results give a more continuous curve of the trimer order parameter when going from the trimer phase to the TSL phase. 

\subsection{Low-energy excitations}

To refine the determination of the phase diagram we have computed the low-energy spectra obtained by ED on a 21-site torus along the selected cuts, as shown in the bottom panels of Figs.~\ref{fig:cut05phi}, \ref{fig:cut00phi} and \ref{fig:cut0125theta}. For a restricted set of parameters, the spectrum has also been computed on a 27-site torus for better accuracy (see inset of Fig.~\ref{fig:cut05phi}).

The spectrum for $(\phi,\theta)=(\pi/2,0)$, on the left of Fig.~\ref{fig:cut05phi}, clearly shows a unique ground state and a gap of order $0.3$ characteristic of the AKLT phase, but the rest of the spectrum seems to come down quickly when increasing $\theta$. We can obtain a complementary estimate of the excitation gap by the MPS excitation ansatz (see section~\ref{appendix:sub_gap} of Appendix~\ref{appendix:MPS}), shown in Fig.~\ref{fig:aklt_gap_imps}, which confirms that the gap decreases very quickly. The right side of Fig.~\ref{fig:cut05phi} shows the finite gap (due to finite size effects) of the fully polarized ferromagnetic phase for $\theta<\pi/3$ (at $\phi=3\pi/2$). Around the pole,  a gapped phase is visible on the 21-site cluster. However, the larger 27-site cluster reveals low-energy (non-singlet) excitations compatible with the magnetic SU(3)-broken phase discussed above.

On the right-hand side of Fig.~\ref{fig:cut00phi}, the even and odd (under inversion) lowest singlets (labeled $\Gamma A$ and $\Gamma B$) are the precursor of the two-fold degenerate trimerized ground state. Between the AKLT and the trimerized region, we see two new low-energy singlets ($\Gamma E_{1a}$) coming down suggesting a three-fold degenerate GS typical of the CSL phase. As discussed before, the small gap seen around the pole is an artifact of the 21-site cluster, no longer present in the larger 27-site cluster.

The ED data in the bottom panel of Fig.~\ref{fig:cut0125theta} are shown on a larger interval along the $\theta=\pi/8$ meridian than the two other panels above, from $\phi=0$ to $\phi=\pi$. It shows the same characteristics encountered in Fig.~\ref{fig:cut00phi} (described in the above paragraph) corresponding to the same sequence (in reverse order) of phases, i.e. the trimer, the TLS, the magnetic and the fully polarized ferromagnet, from left (right) to the right (left) in Fig.~\ref{fig:cut0125theta} (\ref{fig:cut00phi}). Again the trimer (TSL) phase shows two (three) low-energy singlets at the bottom of the spectrum, and a spurious gap appears in the magnetic SU(3)-broken phase (identified by complementary means).

\subsection{Edge modes in the TSL phase}
\label{subsec:chiral_sl}

We shall now discuss further the nature of the topological spin liquid phase on the kagome lattice: our results suggest two candidates, (i) a CSL -- characterized by $\mathbb{Z}_3$ topological order with three sectors -- and (ii) a DCS phase -- characterized by $D(\mathbb{Z}_3)$ topological order with nine sectors. Three different routes have been followed to identify the edge modes: i)~first, by studying the system on an open disk, whose low-energy spectrum should follow that of some $(1+1)d$ CFT; ii)~second, we optimize iMPS on an infinite YC4 cylinder in three different topological sectors, from which the entanglement spectrum can be straightforwardly deduced \cite{CL2013, Bauer2014}; iii)~third, using a faithful TSL representation via symmetric PESS. Note that states with chiral topological order can be faithfully represented by PEPS or PESS \cite{Poilblanc2016, Poilblanc2017b,Hasik2022}, where an infinite correlation length is artificially introduced to generate the chiral features in the entanglement spectrum -- the latter provides a useful diagnostics~\cite{Li2008, Arildsen2022a} for the nature of the TSL.

\begin{figure}[tb]
\includegraphics[width=.95\columnwidth]{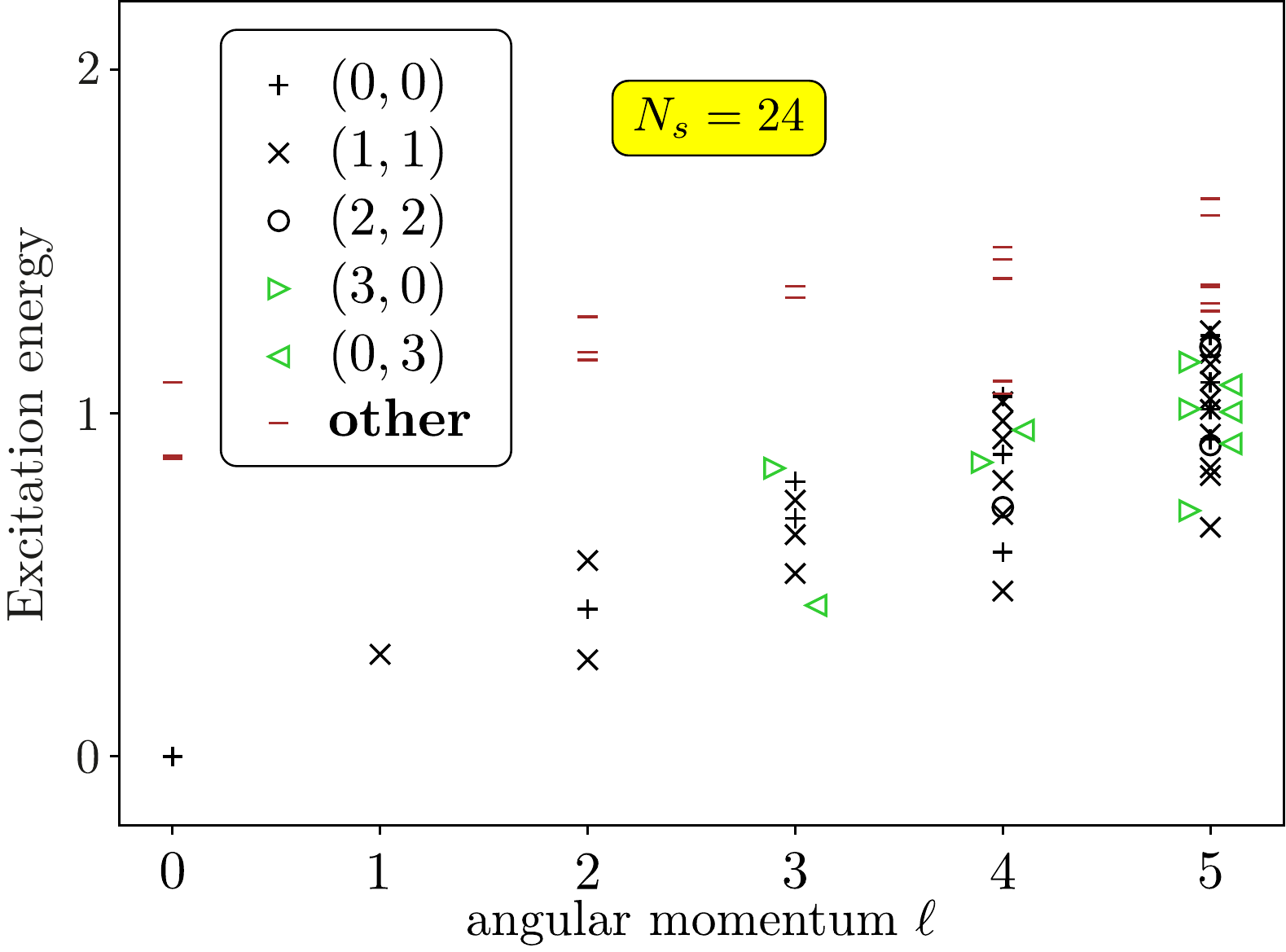}
\caption{
%\justifying
Low-energy spectrum computed with ED on a 24-site kagome cluster with open boundary conditions for $\theta=\pi/4$, $\phi=0$. Relative energies are plotted vs the angular momentum $\ell$ with respect to the C$_6$ rotation symmetry. All symbols agree with the CFT prediction, see Tab.~\ref{tab:wzw-Q0}.
\label{fig:ed_lowenergy_24_obc}
}
\end{figure}
 
Fig.~\ref{fig:ed_lowenergy_24_obc} shows the low energy spectrum in the TSL phase, computed by ED, on a $N_s=24$-site disk. We observe a linearly dispersing chiral mode as a function of the angular momentum associated with the C$_{6}$ symmetry of the cluster. The quantum numbers of the SU(3) multiplets are found to be in good agreement with the WZW SU(3)$_1$ tower of states issued from the singlet ($\bf 1$) ground state (see theoretical expectation in Table~\ref{tab:wzw-Q0}), namely all multiplets are in perfect agreement up to $\ell=3$, while there are few extra multiplets for larger $\ell$ (1 extra {\bf 1} and 1 extra {\bf 8} levels at $\ell=4$, 1 extra {\bf 1}, 2 extra {\bf 8} and 1 extra {\bf 10} levels at $\ell=5$). This small discrepancy could be attributed to the small cluster size. This suggests that the TSL phase is a chiral SL. 

\begin{figure*}[htb]
 \centering
       \includegraphics[width=\textwidth]{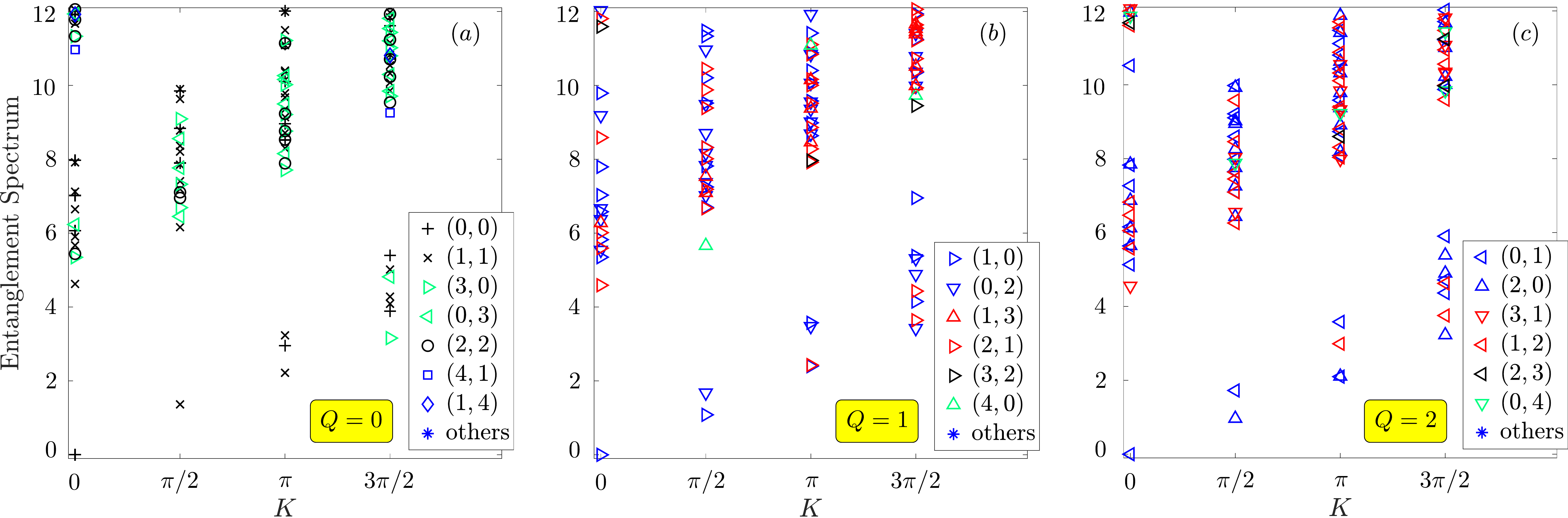}
    \caption{%\justifying
    Entanglement spectra of MPS on a straight $L_y=4$ cylinder (YC4) optimized at $\theta=\pi/4, \phi=0$ in the three topological sectors, with largest MPS bond dimensions around $D_{\mathrm{MPS}}\approx6000$. The three panels correspond to the three topological sectors associated to different total $\mathbb{Z}_3$ charge $Q$ on the boundaries, $Q=0$ $(a)$,  $Q=1$ $(b)$ and $Q=2$ $(c)$. The contents of the chiral modes are consistent with a SU(3) CSL -- see Table~\ref{tab:wzw-Q0} and Table VII of Ref.~\cite{Chen2021}.
    }
    \label{fig:es_mps}
\end{figure*}

\begin{figure*}[htb]
    \centering
    \includegraphics[width=\textwidth]{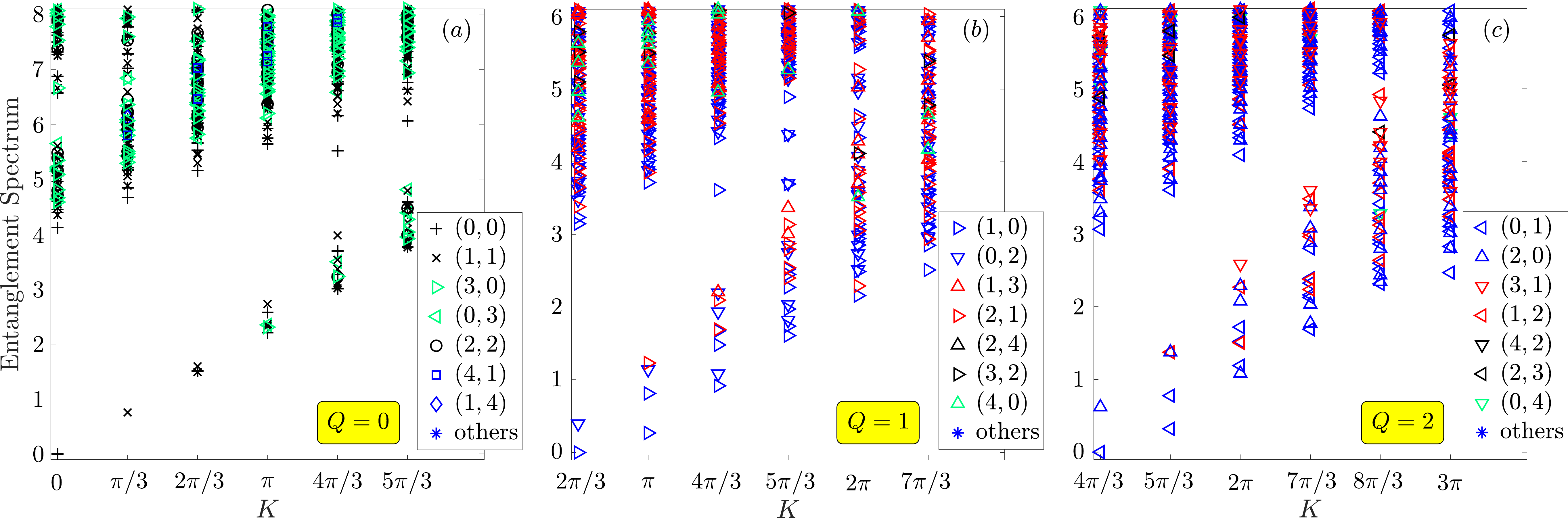}
    \caption{%\justifying
   Entanglement spectra of  a $D=13$ chiral PESS at $\chi=169$ placed on a $L_y=6$ infinite cylinder partitioned in two halves, computed at $\theta=\pi/4, \phi=0$. The virtual space is $\mathcal{V}=\bf{1}\oplus \bf{3} \oplus \bf{\overline{3}} \oplus \bf{6}$.
   The three panels correspond to the three topological sectors associated to different total $\mathbb{Z}_3$ charge $Q$ on the boundaries, $Q=Q({\bf 1})=0$ $(a)$,  $Q=Q({\bf 3})=1$ $(b)$ and $Q=Q({\bf {\overline 3}})=2$ $(c)$.
   The content of the chiral modes agrees with predictions based on the SU(3)$_1$ WZW CFT up to the Virasoro level $L_0=7$ for $Q=0$ (apart from minute deviations) -- see Table~\ref{tab:wzw-Q0} -- and based on the $\mathrm{SU}(3)_1\times \overline{\mathrm{SU}(3)}_1$ DCS theory up to $L_0=4$ otherwise -- see Tables~\ref{tab:wzw-Q1} and \ref{tab:wzw-Q2}.
   Note, in the $Q=0$ sector one 8-dimensional irrep is missing from the compact group of low-energy states of the $L_0=6$ Virasoro level. Also, all relevant irreps of the $L_0=7$ level are grouped together below energy $\sim 7.2$ except three missing (1,1), (0,3), and (2,2) irreps which appear slightly away at energy $\sim 7.32$, $\sim 7.92$ and $\sim 7.52$, respectively.
    \label{fig:es}
    }
\end{figure*}

Similarly, the MPS ES computed on a YC4 infinite cylinder (see Appendix~\ref{appendix:MPS} for further details on the MPS construction) and shown in Fig.~\ref{fig:es_mps} reveal similar features, also supporting the CSL phase. This can be seen in all three sectors corresponding to different $Q$'s. The CFT predictions for these cases can be found e.g. in Tables VI and VII of Ref.~\onlinecite{Chen2021}.

To construct the symmetric PESS, we have followed Ref.~\cite{Kurecic2019} to implement the relevant symmetries in PESS, including $C_3$ rotation symmetry and SU(3) spin rotation symmetry. Moreover, by choosing the appropriate local tensors, the PESS undergoes complex conjugation under reflection, fulfilling the symmetry requirement of both CSL and DCS phases breaking time-reversal symmetry.
One important ingredient in the symmetric PESS construction is the representations carried by the virtual index of local tensors. With $\mathbb{Z}_3$ gauge symmetry in mind, a minimal virtual space would be ${\cal V}={\bf 1}+{\bf 3}+\overline{\bf 3}$, which was shown to support a $\mathbb{Z}_3$ toric code type topological phase in the parameter space. It turns out, doing variational optimization in this class of PESS always runs into a simplex solid phase, where the up and down triangles become inequivalent. This could be understood from spontaneous symmetry breaking at the PESS level. Therefore, one has to consider a larger virtual space for representing SU(3) CSL phase. For that, we have used a SU$(3)$ symmetric simple update algorithm (implemented with the QSpace library~\cite{Weichselbaum2012, Weichselbaum2020}) combined with variational optimization, and found that virtual irreps ${\cal V}={\bf 1}+{\bf 3}+\overline{\bf 3}+{\bf 6}$ and ${\cal V}={\bf 1}+{\bf 3}+\overline{\bf 3}+{\bf 6}+{\bf 8}$ could provide a good description of the TSL.

The entanglement spectrum (ES) with virtual space ${\cal V}={\bf 1}+{\bf 3}+\overline{\bf 3}+{\bf 6}$ is computed and shown in Fig.~\ref{fig:es}. Using the $\mathbb{Z}_3$ gauge symmetry, we group the ES into three sectors with $\mathbb{Z}_3$ charge $Q=0,1,2$, respectively. The momentum $K$ around the circumference of the cylinder is a good quantum number and is used to label the ES.

As shown in Fig.~\ref{fig:es}, we have obtained linear dispersing chiral branches in the low energy part of the ES in all three sectors. A close look at the content of the  chiral branch in the $Q=0$ sector  compared to predictions of the WZW SU(3)$_1$ CFT in Table~\ref{tab:wzw-Q0} reveals that the relevant SU$(3)$ multiplets were captured up to the 7th Virazoro level apart from minute deviations (see figure caption). 

However, zooming in at the level content of the $Q=1$ and $Q=2$ sectors, one finds that the quasi-degenerate SU$(3)$ multiplet structure differs from the simple tower of states given by the WZW CFT. Instead, the low energy spectrum in the $Q=1$ sector can be explained by the tensor product of $\bf{\overline{3}}$ with the $Q=2$ SU(3)$_1$ CFT tower. Similar considerations apply to the $Q=2$ sector giving the conjugate irreps of the $Q=1$ sector. A comparison with Tables~\ref{tab:wzw-Q1} and \ref{tab:wzw-Q2} shows that the counting is indeed consistent with 
${\overline{3}}$-tower [$\otimes \overline{3}$] and its conjugate, respectively, up to the 4th Virasoro level (for $Q=1$, at $L_0=3$ one $\bf{\overline 6}$ irrep lies a bit away from the compact group of the remaining SU(3)$_1$ irreps). Similar features have been found in a simpler PESS on the kagome lattice with virtual space ${\cal V}={\bf 1}+{\bf 3}+\overline{\bf 3}$~\cite{Kurecic2019}. It was further established that this PESS belongs to a $\mathrm{SU}(3)_1\times \overline{\mathrm{SU}(3)}_1$ double Chern-Simons phase characterized by a slow SU$(3)_1$ chiral mode and a fast counter-propagating $\overline{\mathrm{SU}(3)}_1$ chiral mode~\cite{Arildsen2022b}.  Our findings suggest that our $D^*=4$ ($D=13$) PEPS ansatz also belongs to the same phase. However, it is unclear whether the presence of a second fast mode is a genuine feature of the phase or a finite-$D$ effect. Note that the ED results do not show evidence of the DCS phase: for instance, in Fig.~\ref{fig:cut00phi} a 3-fold quasi-degenerate ground state manifold is seen on the torus around $(\phi,\theta)=(0,\pi/4)$, in agreement with the expectation for a chiral SL (while a 9-fold degenerate ground state is expected in a DCS phase). Similarly, the ES obtained in the MPS simulations in 3 topological sectors (see Appendix~\ref{appendix:sub_csl} for details) differ from the ones obtained in the PEPS framework, as shown in Fig~\ref{fig:es_mps}, being compatible with a SU(3)$_1$ CSL. 

It would be interesting to analyze whether the level splittings in ES can be described by a generalized Gibbs ensemble~\cite{Arildsen2022b}. Very recently~\cite{Arildsen2023} such a refined analysis of the ES enabled to strengthen the evidence for the CSL nature of a SU(3) PEPS on the square lattice~\cite{Chen2020}. In that case, it was shown that the splittings between conjugate irreps in the same Virasoro level of $Q=0$ sector and between $Q=1$ and $Q=2$ sectors should vanish (and numerically found to be extremely small compared to the scale of other level splittings), due to absence of certain conserved quantities which are not allowed by symmetry. In the present case (Fig.~\ref{fig:es}), the splittings between conjugate irreps $(3,0)$ and $(0,3)$ at $L_0=4,5$ in the $Q=0$ sector is noticeable, and the entanglement energies between conjugate irreps in the $Q=1$ and $Q=2$ sectors also have small but visible differences. On the other hand, the entanglement spectrum from MPS calculation, shown in Fig.~\ref{fig:es_mps}, agrees with the level counting of a chiral SL, but also has a splitting between conjugate irreps at the same Virasoro level (in $Q=0$ sector or between $Q=1$ and $Q=2$ sectors). A full analysis of the splittings in Fig.~\ref{fig:es_mps} and Fig.~\ref{fig:es} is left for future work.

%%%%%%%%%%%%%%%%%%
\section{Conclusions} %
%%%%%%%%%%%%%%%%%%
\label{sec:conclusion}

In this work, the investigation of the complete phase diagram of the SU(3) chiral Heisenberg model on the kagome lattice has been carried out. Physical spins transforming according to the fundamental irrep of SU(3) are considered on all lattice sites. The Hamiltonian includes the generic two-site and three-site couplings on nearest-neighbor sites and triangular units, respectively. To map out the phase diagram we have combined analytical and numerical tools such as magnon expansions, exact diagonalisations and tensor network (iMPS and iPEPS) techniques. In addition to the AKLT phase predicted by one of the authors (KP)~\cite{Penc2022} and the (expected) fully polarized ferromagnet (at large enough ferromagnetic couplings) we have found two gapped singlet phases and a magnetic phase that spontaneously breaks the SU(3) symmetry. 

One of the singlet phases, the trimer phase, breaks spontaneously the inversion center exchanging up and down triangles, as observed in the pure SU(3) Heisenberg model~\cite{Corboz2012b}, a special point in our 2D parameter space. We have also found an enigmatic topological spin liquid. Although our numerical results show evidence for a gap and $\mathbb{Z}_3$ gauge symmetry (via MPS and PEPS constructions), the exact nature of this TSL phase is still controversial with two possible candidates, either a SU(3)$_1$ CSL (proposed in Ref.~\cite{WuTu2016}) or a double $\mathrm{SU}(3)_1\times \overbar{\mathrm{SU}(3)_1}$ Chern-Simons spin liquid (discussed in Refs.~\cite{Kurecic2019,Arildsen2022b}). 

The not fully polarized SU(3)-broken magnetic phase is, most likely, a uniform partially polarized ferromagnet with collinear spins in the 3-site unit cell. Another competing non-uniform phase  with spin canting occurring on three triangular sub-lattices separately (requiring a 9-site unit cell) seems to be slightly disfavored energetically.

%%%%%%%%%%%%%%%%%%
\smallskip

\par\noindent\emph{\textbf{Acknowledgments}} --- %
%%%%%%%%%%%%%%%%%%
We acknowledge the participation of Seydou-Samba Diop (Ecole Normale Sup\'erieure de Lyon) and Matthew W. Butcher (Rice University) at an early stage of the project and support from the TNTOP ANR-18-CE30-0026-01 grant awarded by the French Research Council, and the European Research Council (ERC) under the European Union's Horizon 2020 research and innovation programme (grant agreement No 101001604). This work was granted access to the HPC resources of CALMIP center under the allocations 2022-P1231 and 2022-P0677 as well as GENCI (grant x2021050225). We also acknowledge Jan von Delft for providing part of the computational resource. J.-Y.C. was supported by Open Research Fund Program of the State Key Laboratory of Low-Dimensional Quantum Physics (project No.~KF202207), Fundamental Research Funds for the Central Universities, Sun Yat-sen University (project No.~23qnpy60), a startup fund from Sun Yat-sen University (No.~74130-12230034), and the Innovation Program for Quantum Science and Technology 2021ZD0302100. L.V. is supported by the Research Foundation Flanders. Y.X. and A.H.N. were supported by the Division of Materials Research of U.S. National Science Foundation under the Award DMR-1917511. The iPESS and iPEPS calculations at Rice University were supported in part by the Big-Data Private-Cloud Research Cyber infrastructure MRI-award funded by NSF under grant CNS-1338099 and by Rice University's Center for Research Computing (CRC). K.P. acknowledges support from the Hungarian NKFIH OTKA Grant No.  K142652.

\clearpage

\appendix

\section{Analytical developments}
\label{appendix:analytic}

\subsection{Stereographic projection}
\label{appendix:analytic-stereo}

\begin{figure}[htb]
	\centering
		\includegraphics[width=\columnwidth]{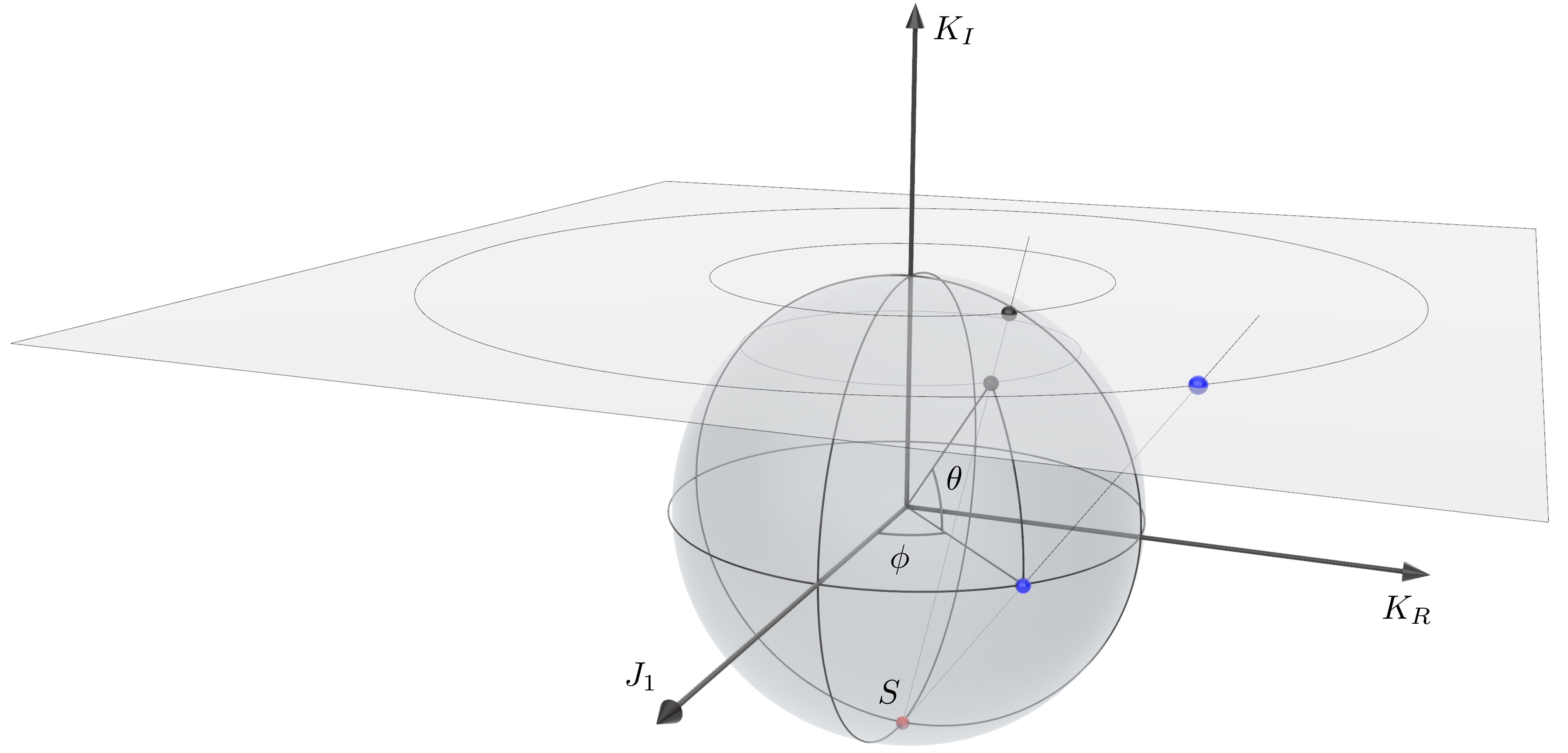}
	\caption{%\justifying
 Stereographic projection of the North hemisphere from the South pole, mapping every point of the sphere such as $0\leq\theta\leq\pi/2$ and $0\leq\phi<2\pi$ to its image on the upper planar disk.
		\label{fig:stereo}}
\end{figure}

The stereographic projection (see Fig.~\ref{fig:stereo}) maps the parameter space (see Eq.~(\ref{eq:paras}) for $0\leq\theta\leq\pi/2$ and $0\leq\phi<2\pi$ to a planar disk delimited by the image of the equator (a circle of radius 2). The image coordinates of $(\theta,\phi)$ points on the projection plane are given by
\begin{eqnarray}
	X&=&\frac{\cos (\theta ) \cos (\phi )}{\sin (\theta )+1}, \nonumber\\
	Y&=&\frac{\cos (\theta ) \sin (\phi )}{\sin (\theta )+1}.\nonumber
\end{eqnarray}

\begin{table}[!ht]
    \centering
    \begin{tabular}{p{.2\columnwidth} p{.3\columnwidth} p{.3\columnwidth}} \hline \hline 
    Symbols & Dynkin labels  &  Young tableaus \\
    \hline
    \includegraphics[width=3mm]{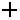} &(0,0) & $\Yvcentermath1\su{0}{0}{1}$  \\
    \includegraphics[width=3mm]{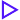} &(1,0) & $\Yvcentermath1\su{0}{1}{3}$ \\
    \includegraphics[width=3mm]{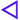} &(0,1) & $\Yvcentermath1\su{0}{1,1}{\bar{3}}$\\
    \includegraphics[width=3mm]{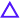} &(2,0) & $\Yvcentermath1\su{0}{2}{6}$ \\
    \includegraphics[width=3mm]{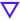} &(0,2) & $\Yvcentermath1\su{0}{2,2}{\bar{6}}$  \\
    \includegraphics[width=3mm]{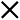} &(1,1) & $\Yvcentermath1\su{0}{2,1}{8}$\\
    \includegraphics[width=3mm]{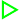} &(3,0) & $\Yvcentermath1\su{0}{3}{10}$\\
    \includegraphics[width=3mm]{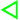} &(0,3) & $\Yvcentermath1\su{0}{3,3}{\overline{10}}$ \\
    \includegraphics[width=3mm]{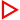} &(2,1) & $\Yvcentermath1\su{0}{3,1}{15}$\\
    \includegraphics[width=3mm]{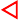} &(1,2) & $\Yvcentermath1\su{0}{3,2}{\overline{15}}$ \\
    \includegraphics[width=3mm]{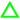} &(4,0) & $\Yvcentermath1\su{0}{4}{15'}$\\
    \includegraphics[width=3mm]{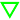} &(0,4) & $\Yvcentermath1\su{0}{4,4}{\overline{15}'}$ \\
      \includegraphics[width=3mm]{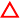} &(1,3) & $\Yvcentermath1\su{0}{4,3}{24}$\\
     \includegraphics[width=3mm]{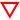} &(3,1) & $\Yvcentermath1\su{0}{4,1}{\overline{24}}$\\
    \includegraphics[width=3mm]{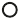} &(2,2) & $\Yvcentermath1\su{0}{4,2}{27}$ \\   
    \includegraphics[width=3mm]{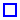} &(4,1) & $\Yvcentermath1\su{0}{5,1}{35}$ \\
    \includegraphics[width=3mm]{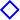} &(1,4) & $\Yvcentermath1\su{0}{5,4}{\overline{35}}$ \\
    \includegraphics[width=3mm]{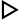} &(3,2) & $\Yvcentermath1\su{0}{5,2}{42}$ \\
    \includegraphics[width=3mm]{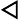} &(2,3) & $\Yvcentermath1\su{0}{5,3}{\overline{42}}$ \\
    \includegraphics[width=3mm]{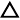} &(2,4) & $\Yvcentermath1\su{0}{6,4}{60}$ \\
       \includegraphics[width=3mm]{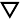} &(4,2) & $\Yvcentermath1\su{0}{6,2}{\overline{60}}$\\    \includegraphics[width=3mm]{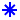} &\multicolumn{2}{c}{Others}\\
    \hline \hline
    \end{tabular}
    \caption{%\justifying
    Correspondence between Dynkin labels and Young tableaus for all SU(3) irreps discussed in the paper. Symbols used in the various figures are displayed in the first column. Labeling of the irreps follows conventions from the LieArt package~\cite{Feger2020}.
    }
    \label{tab:tensors}
\end{table}

\subsection{SU(3) irreducible representations and conformal towers}
\label{appendix:analytic-su3}

Irreducible representations (irreps) of the SU(3) group can be labeled differently. We show in Table~\ref{tab:tensors} the one-to-one correspondence between Dynkin labels and Young tableaus. 
An irrep labeled by $(x,y)$ corresponds to a Young tableau with two rows containing $x+y$ and $y$ boxes.
Conformal towers of various WZW CFTs mentioned in the main text, originating from $\bf{1}$, $\bf{\overline{3}}$ [$\otimes\bf{\overline{3}}$] and $\bf{3}$ [$\otimes\bf{3}$], are shown in Tables~\ref{tab:wzw-Q0},\ref{tab:wzw-Q1} and \ref{tab:wzw-Q2}, respectively. See also Ref.~\cite{Chen2021} for the towers originating from $\bf{3}$ and $\bf{\overline{3}}$.

%\clearpage
%\section{One- and two-magnon spectra}

\subsection{Single magnon dispersion}
\label{appendix:analytic-magnon}

 The dispersion of a single magnon is determined by the eigenvalues of the matrix in reciprocal space given in Eq.~\ref{eq:magnon} below, where the energy is measured from the energy of the ferromagnetic state and $\mathbf{q}=(q_x,q_y)$ is the momentum of the magnon. 
  The $J+K_R$ appear together in the matrix above, so the dispersion depends only on two free parameters $J+K_R$ and $K_I$.  
 \clearpage
\begin{widetext}
\begin{equation}
\left(
\begin{array}{ccc}
 -4 (J+K_R) 
 & 2 (J + K_R -i K_I) \cos \frac{q_x - \sqrt{3} q_y}{2} 
 & 2 (J + K_R +i K_I) \cos \frac{q_x + \sqrt{3} q_y}{2} 
 \\
 2 (J + K_R +i K_I) \cos \frac{q_x - \sqrt{3} q_y}{2}
 & -4 (J+K_R) 
 & 2 (J + K_R - i K_I) \cos q_x
 \\
 2 (J + K_R -i K_I) \cos \frac{q_x + \sqrt{3} q_y}{2}
 & 2  (J + K_R +i K_I) \cos q_x
 & -4 (J+K_R) 
 \\
\end{array}
\right)
\label{eq:magnon}
\end{equation}
\end{widetext}

$N-1$ states belong to the $\mathbf{d}=(N-1)(N+1)$ dimensional Young diagram with $(N-2,1)$ Dynkin label, and the state with zero energy and $\mathbf{q}=0$ to the $\mathbf{d}=(N+1)(N+2)/2$ dimensional fully symmetrical irreducible representation of the ferromagnetic state, represented by the Young diagram $(N,0)$.

\subsection{Two-magnon spectra}
\label{appendix:analytic-2magnons}

\begin{figure}[hb]
\includegraphics[width=.9\columnwidth]{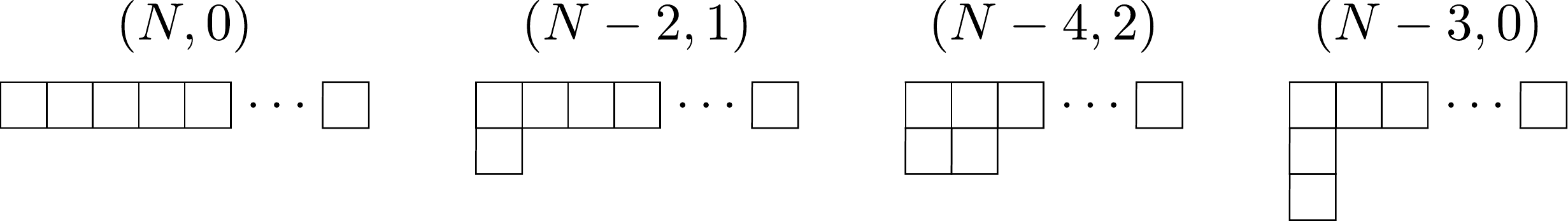}
\caption{%\justifying
The Young diagrams appearing in the $[N_A,N_B,N_C] = [N-2,1,1]$ sector of the two magnon calculations, labeled by their Dynkin indices. $N_A$ is the number of sites having $A$ spins, and so on, so that $N_A+N_B+N_C=N$.  \label{fig:young_diagrams_2mag}
}
\end{figure}

\begin{figure}[tb]
\includegraphics[width=.9\columnwidth]{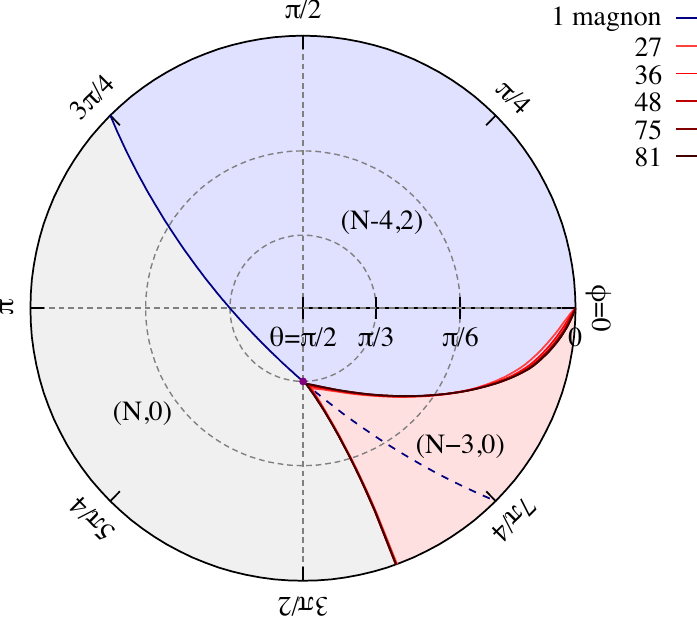}
\caption{
%\justifying
The ground state diagram in the two-magnon sector. The blue line denotes the 1-magnon instability line. The two magnons added to the system can form a symmetrical [Young diagram with Dynkin label $(N-4,2)$, blue shaded region] or antisymmetrical [Young diagram $(N-3,0)$, red shaded region] combination; the regions indicate regions where they are the lowest energy states. The boundaries for system sizes from 27 to 81 are drawn in red lines.
 \label{fig:2magnon_phase_diagram}
}
\end{figure}

This section considers two magnon excitations of the fully-symmetrical (ferromagnetic) state, where we introduce two spins with different flavors, following the calculation of Refs.~\onlinecite{Wortis1963, Hanus1963} for the SU(2) case. Starting from the $|AA\dots A\rangle$ as a vacuum, the two-magnon wave function is 
\begin{equation}
\Psi = \sum_{i,j\in \Lambda} c_{i,j} |A\dots AB_iA \dots AC_jA \dots A\rangle\;.
\end{equation}
The dimension of the Hilbert space spanned by $|A\dots AB_iA \dots AC_jA \dots A\rangle$ basis is $N (N-1)$, as $i = 1,\dots,N$ and $j = 1,\dots,N$, but the $B$ and $C$ cannot occupy the same site ($i\neq j$). Furthermore, we symmetrize and antisymmetrize the wave functions so that 
\begin{align}
 c^e_{i,j} &= c_{i,j} + c_{i,j} \,,
\nonumber\\
 c^o_{i,j} &= c_{i,j} - c_{i,j} \,.
\nonumber
\end{align}
The dimensions of the symmetric ``e" (even) and of the antisymmetric ``o" (odd) subspace are the same and equal to $N(N-1)/2$. The even subspace is composed of the $(N,0)$, $(N-2,1)$, and the $(N-4,2)$ Young diagrams (see Fig.~\ref{fig:young_diagrams_2mag}), each having multiplicities $1$, $N-1$, and $N(N-3)/2$, respectively. The irreducible representations in the odd subspace are $(N-2,1)$ and the $(N-3,0)$ Young diagrams with multiplicities $N-1$ and $(N-2) (N-1)/2$. Using the Casimir operator, we separate the energies of the $(N-4,2)$ and  $(N-3,0)$ irreducible representations. The symmetric, even sector would also appear in the SU(2) case since symmetrization is equivalent to taking two $B$ type spins instead of $B$ and $C$. The odd (antisymmetric) sector is unique to the SU(3).
We diagonalized the Hamiltonian matrix for up to 81-site clusters numerically. Since this is a two-body problem, one might derive analytic expressions in principle, but they would be quite cumbersome.

\begin{figure}[hb]
\includegraphics[width=\columnwidth]{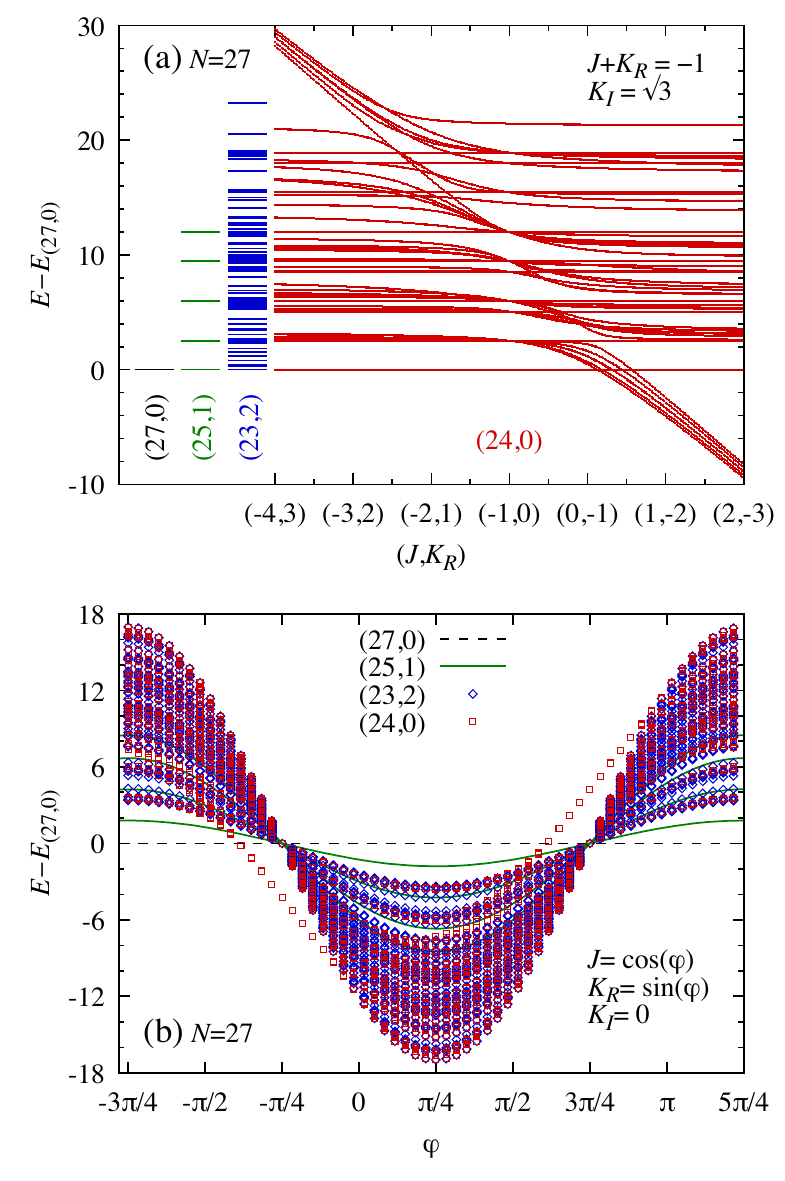}
\caption{
%\justifying
The one- and two-magnon spectra for the 27-site cluster relative to the energy of the ferromagnetic state with $(27,0)$ Young diagram. Shown are the energies of the one-magnon states [green, (25,1) Young diagram] and the symmetric [blue,(25,1)] and antisymmetric [red,(24,0)] two-magnon states.
(a) Varying the $J$ and $K$ while keeping the $J+K_R=K_I/\sqrt{3} = -1$ constant follows the one-magnon instability line. The energies of the lowest unbound states are equal; for positive values of $J$, 18 two-magnon bound states detach from the continuum (red lines with $E_{(24,0)} < E_{(27,0)}$). 
(b) The one- (continuous green curves) and two-magnon (open symbols) energies for $K_I=0$ (i.e. $\theta=0$). The two-magnon bound states appear for $-\pi/2<\phi<0$.
\label{fig:enes_2magnon_multi}
}
\end{figure}

 In Sec.~\ref{sec:ferro_main}, we derived the conditions for the stability of one-magnon excitations. When the energy of the one-magnon is larger than that of the ferromagnetic state, the ferromagnetic phase is the ground state. If the energy needed to create a magnon is negative, the ferromagnetic phase is not a ground state anymore. The importance of the two magnon calculation is to reveal the interaction between the magnons. The ferromagnetic phase shrinks if the magnons attract each other and form bound states.  

Fig.~\ref{fig:2magnon_phase_diagram} summarizes the result of our calculation. It shows the Young diagram (YD) having the lowest energy for two magnons. We can distinguish three regions: the $(N,0)$ ferromagnetic phase (the gray area), the $(N-4,2)$ for the symmetric combination of the two magnons (the blue area), and the red-colored area where the antisymmetric combination of $(N-3,0)$ Young diagram is the ground state. The boundary between the $(N,0)$ and the $(N-4,2)$ follows the one-magnon instability line for negative values of $J$, but at $\theta=\pi/3$ and $\phi=3\pi/2$ corresponding to $J=0$, $K_R=-1$ and $K_I = \sqrt{3}$ the three regions meet, and the $(N-3,0)$ antisymmetric combination becomes the ground state. The boundary between the ferromagnetic phase and the  $(N-3,0)$ no longer follows the one-magnon instability line. It is a hint for a formation of a bound state. 

\begin{figure}[tb]
\includegraphics[width=\columnwidth]{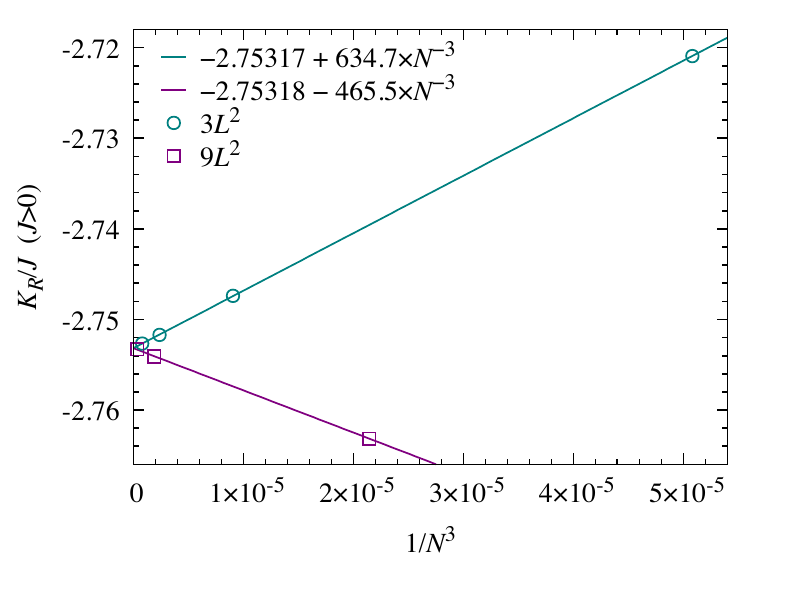}
\caption{
%\justifying
The finite size scaling of the boundary between the $(N,0)$ ferromagnetic state and the $(N-3,0)$ state for $K_I=0$. The boundary for the $3L^2$ and $9L^2$ type clusters goes to the same $K_R/J = -2.7532$ value in the thermodynamic limit, though the slopes in $1/N^3$ are quite different.\label{fig:FM_sing_bnd}}
\end{figure}

\begin{figure}[tb]
\includegraphics[width=.9\columnwidth]{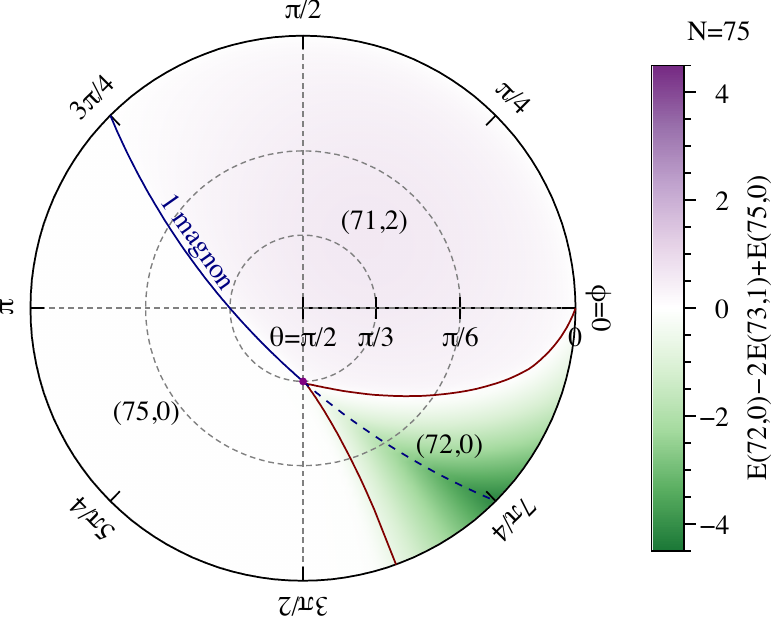}
\caption{
%\justifying
The value of the energy gap $E(N-3,0) - 2 E(N-2,1)+E(N,0)$ for the $75$-site cluster.
The bound state appears when the gap is negative; this is the green area in the plot.
\label{fig:2magnon_gap}
}
\end{figure}

 To get further insight, we plot in Fig.~\ref{fig:enes_2magnon_multi}(a) the energies of the different irreducible representations along the $J+K_R = \pm \sqrt{3} K_I$ one-magnon instability line for a 27 site cluster. The lowest energies of the ferro state $(N,0)$, the one magnon $(N-2,1)$, and the two magnons in a symmetric combination $(N-4,2)$ are all equal (we note that the energies in these irreducible representations depend on the $J+K_R$ combination only). In the  $(N-3,0)$ antisymmetric sector, a band of bound-states appears with lower energy for $J \gtrsim 0$ in the figure, where we keep $J + K_R = -1$ constant (in the thermodynamic limit, the triple point is at $J=0$, $K_R=-1$, and $K_I=\sqrt{3}$, i.e. $\theta=\pi/3$ and $\phi=3\pi/2$). The number of bound-states is 18, which is equal to the number of triangles in the 27-site kagome cluster. We also confirmed that the number of bound states is $2N/3$ in other clusters. In Fig.~\ref{fig:enes_2magnon_multi}(b), we plot the energy gap to the ferromagnetic state around the full circle, keeping $K_I = 0$ (i.e., $\theta=0$). At the special point $\phi=-\pi/4$, which corresponds to $J = -K_R$ with $J$ positive, the spectrum of the $(N-3,0)$ states greatly simplifies: we get a $2N/3$-fold degenerate manifold at $E=-6 J $, and all the other energy levels collapse at $E=0$. The explanation is simple: the $B$, $C$ and an $A$ form a localized SU(3) singlet on a triangle, while the remaining $N-3$ $A$ spins constitute the ferromagnetic background.  The singlets can form on any of the $2N/3$ elementary triangles in the kagome lattice, and this is the origin of the degeneracy.  The result of a finite-size scaling for the boundary between the ferromagnetic state and the $(N-3,0)$ state is presented in Fig.~\ref{fig:FM_sing_bnd} for $K_I=0$. We get $K_R/J=-2.7532$, which corresponds to $\theta = 1.611\pi$.
 Figure~\ref{fig:2magnon_gap} shows the energy gap in the full parameter space.
 In this region where the gap is finite, the ground state in the $N_A=N_B=N_C = N/3$ sector is the trimerized state. We can think of it as a condensation of the local SU(3) singlets with a repulsive interaction between them. 

The dispersion of a single magnon in the ferromagnetic background is flat along the one-magnon instability line. The flat bands are connected with modes localized on hexagons, also with delocalized modes, since a dispersing band touches the flat band. When we diagonalize the two-magnon spectrum along the instability line $J+K_R = \pm \sqrt{3} K_I$, the number of states degenerate with the ferromagnet is $\binom{N_{\text{hex}}-1}{2}$ for the symmetric $\text{YD}=(N-4,2)$ and $\binom{N_{\text{hex}}}{2}$ for the antisymmetric  $\text{YD}=(N-3,0)$ combination, where $N_{\text{hex}} = N/3$ is the number of hexagons. 

\section{Lanczos Exact Diagonalization}
\label{appendix:ED}
\begin{figure}[htb]
	\centering
		\includegraphics[width=\columnwidth]{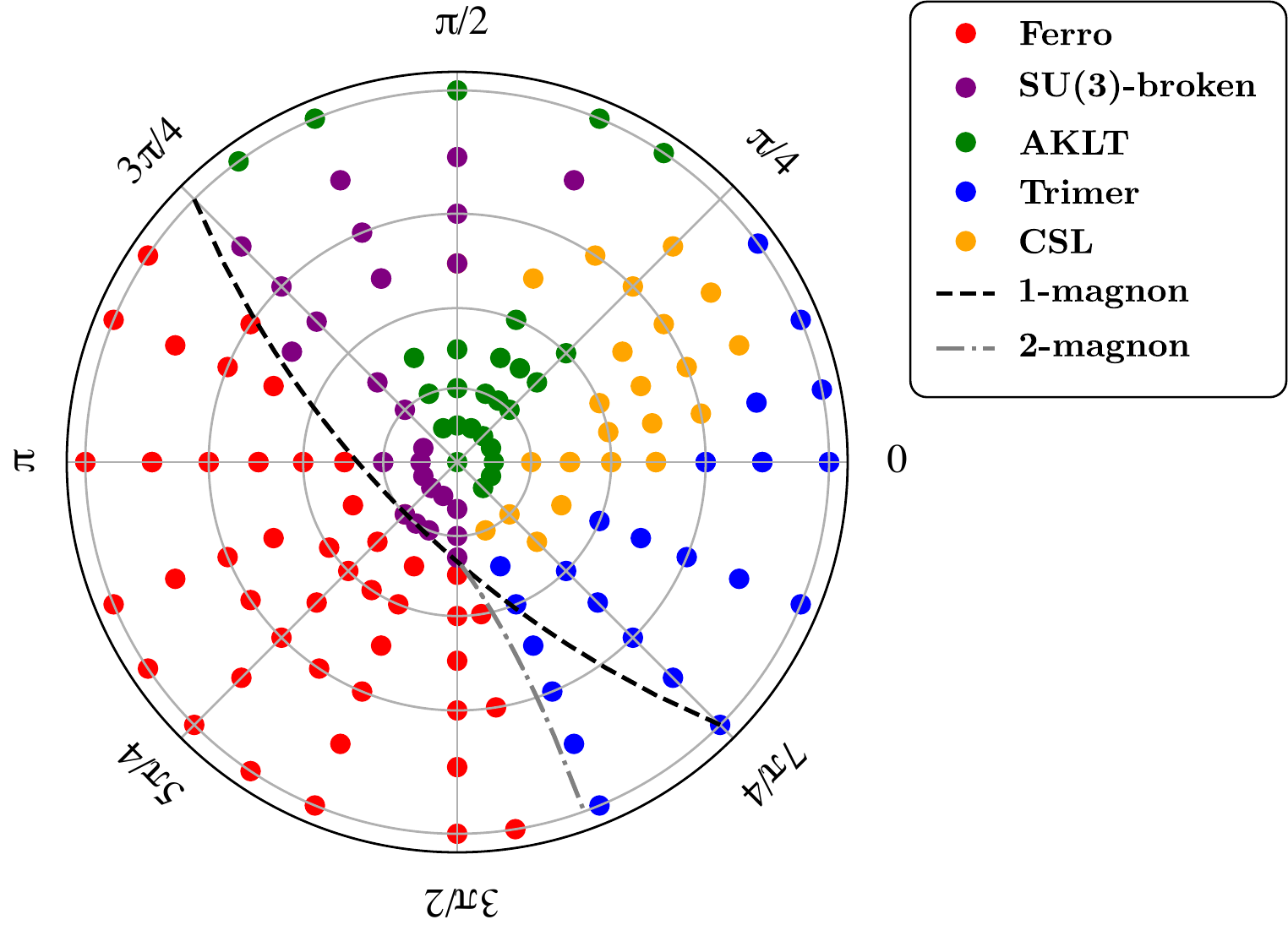}
	\caption{
    %\justifying
    Stereographic projection (same parametrisation as in Fig.~\ref{fig:phasediag}) of the phase diagram of the SU(3) chiral antiferromagnet on the kagome lattice obtained from ED on a 21-site periodic cluster.
		The various phases discussed in the text are represented by dots of different colors. The dashed (dash-dotted) line corresponds to the single (two) magnon instability of the ferromagnetic phase. 
		\label{fig:phasediag2}
}
\end{figure}

We have performed ED on various lattices using space group symmetries as well as color conservation (equivalent to the conservation of the 2 U(1) Cartan generators of SU(3)). By using a Lanczos algorithm, we are able to converge a few low-energy states in each symmetry sector. A typical energy plot is shown for instance in the bottom panel of Fig.~\ref{fig:cut05phi}.

In order to sketch a tentative phase diagram using a single system size, we have computed systematically the low-energy spectrum on the 21-site kagome cluster with periodic boundary conditions on a fine $(\phi,\theta)$ parameter grid. For each set of parameters, we have attempted in Fig.~\ref{fig:phasediag2} to determine its ground-state (GS) properties using the following criteria:
\begin{itemize}
\item ferromagnetic phase: the finite-size GS belongs to the fully symmetric irrep and its energy is known exactly.
\item AKLT: the ground-state is non-degenerate and there is an apparent large gap to the first excitation.
\item Trimerized: there are two low-energy singlet states in the $\Gamma$.A and $\Gamma$.B irreps, as expected if the inversion symmetry is broken in the thermodynamic limit.
\item CSL: there are three low-energy singlet states at momentum $\Gamma$ as expected for a chiral spin liquid on a torus.
\item SU(3)-broken: either the GS is not an SU(3) singlet, or there is a small gap to a non-singlet state. This could be a critical state or a canted ferromagnet.
\end{itemize}
By using these rules, we are able to plot a qualitative phase diagram in Fig.~\ref{fig:phasediag2}.

Note that finite-size effects have been shown to be important in some regions. For instance, our ED data on the 21-site cluster are rather similar at the exact AKLT point $(\phi=\pi/2,\theta=0)$ and close to the North pole (see Fig.~\ref{fig:cut05phi}) so that both regions are labelled in the same way on the phase diagram in Fig.~\ref{fig:phasediag2}. However, the situation is radically different on the 27-site cluster (see inset of Fig.~\ref{fig:cut05phi}), which rather indicates an SU(3)-broken phase in a large region around the North pole. Hence, it is crucial to combine different numerical methods in order to get reliable results.

\section{Cylinder MPS simulations}
\label{appendix:MPS}

\subsection{Geometry}
\begin{figure} \begin{center}
\includegraphics[width=0.95\columnwidth]{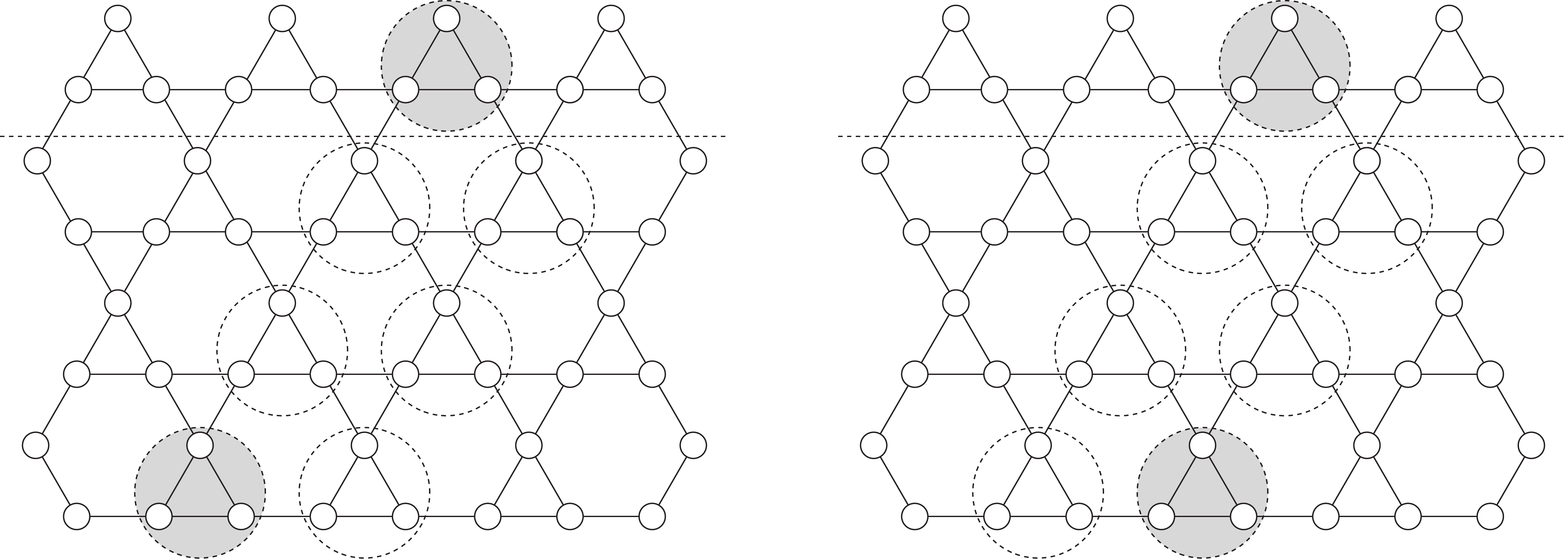}
\end{center} 
\caption{
%\justifying
Different geometries for the kagome lattice on an infinite cylinder with a three-triangle unit cell in the periodic direction; the grey-shaded triangles are identified.}
\label{fig:kagome}
\end{figure}

There are different possible geometries possible for putting the kagome lattice on an infinite cylinder. The most common one is the YC cylinder, shown in Fig.~\ref{fig:kagome}(a). Here we choose a slightly different geometry, shown in Fig.~\ref{fig:kagome}(b), where we have shifted the periodic boundary conditions with a single triangle in the horizontal direction. This shifted geometry has the advantage that the unit cell of the resulting one-dimensional Hamiltonian has a single-triangle unit cell.

\subsection{MPS ground states}

The shifted geometry has the additional advantage that also the MPS ground-state approximation can be chosen to have a single-triangle unit cell. We can represent this MPS as
\begin{equation}
\ket{\Psi(A)} = \mydiagram{3},
\end{equation}
where for convenience we have grouped the three physical $\mathbf{3}$ spins in a single MPS tensor -- in the simulations we always keep three different MPS tensors. We impose $\SU(3)$ symmetry on the MPS, which implies that the virtual degrees of freedom in the MPS can be labeled by $\SU(3)$ irreps. The different blocks in the MPS tensor need to obey the $\SU(3)$ fusion rules, i.e. we need virtual irreps $I_v$ and $I_v'$
\begin{equation}
\mydiagram{4}.
\end{equation}
Now we use the $\Z_3$ property of the $\SU(3)$ fusion rules, where we can group the irreps in three different groups with a $\Z_3$ charge:
\begin{equation}
\begin{cases}
 \overline{\mathbf{3}}, \mathbf{6}, \dots : Q=-1 \\
 \mathbf{1}, \mathbf{8}, \dots : Q=0 \\
 \mathbf{3}, \overline{\mathbf{6}}, \dots : Q=+1 \\
 \end{cases}
\end{equation}
The three physical spins transform jointly as $Q=0$ irreps, so the $\Z_3$ property of the fusion rules dictates that $I_v$ and $I_v'$ can only contain irreps from one and the same group, and, therefore, that we have three classes of MPS labeled by the $\Z_3$ charge of the irreps on the bonds. Depending on the phase we are simulating, the optimal iMPS ground state will be found in one or more of these classes. 

The diagnostics for deciding whether an optimal iMPS is found within a certain $\Z_3$ sector is through the entanglement spectrum and transfer matrix spectrum. In the following, we illustrate this procedure for the different SU(3) singlet phases in the phase diagram. In Figs.~\ref{fig:comb_aklt}, \ref{fig:comb_trimer} and \ref{fig:comb_csl}, we plot the entanglement spectrum and transfer matrix spectrum for three different parameter choices, each time with iMPS optimized in the three different classes. For the entanglement spectrum 
 we plot the different entanglement eigenvalues or Schmidt values with magnitude on the axis; the different colors in Figs.~\ref{fig:comb_aklt}-\ref{fig:comb_csl} correspond to the different SU(3) quantum numbers, which are labeled on the horizontal axis by their Dynkin label. For the transfer matrix spectrum (unit circles in the complex plane exhibited in the insets), we show a few dominant eigenvalues in the first two SU(3) sectors (again denoted by their Dynkin label) in the complex plane. 

\subsection{AKLT phase}

Let us first consider the exact AKLT state, which is represented as a PESS with virtual irrep $\overline{\mathbf{3}}$. On an infinite cylinder, this state can be represented as a snake-like iMPS by dragging the virtual links along the MPS around the cylinder. The virtual links of the iMPS then contain a number of $\overline{\mathbf{3}}$ irreps that scales with the circumference of the cylinder, which are then fused into a number of SU(3) irreps on the virtual leg of the iMPS. Therefore, the $\Z_3$ quantum number of the MPS depends on the cylinder circumference.
\par
\begin{figure} \begin{center}
		\includegraphics[width=\columnwidth]{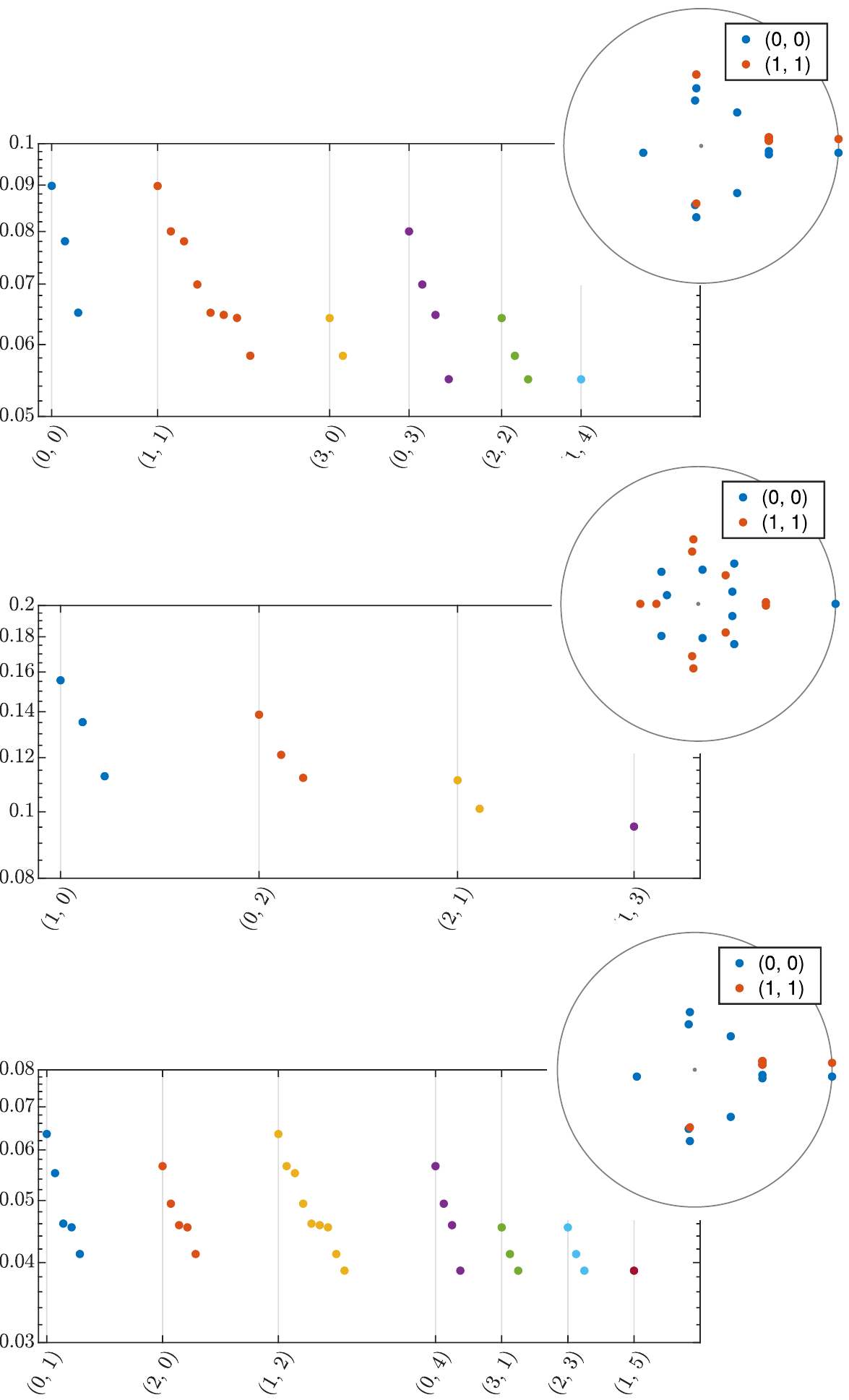}
\end{center} 
\caption{
%\justifying
MPS entanglement spectra of the AKLT state ($\theta=0$, $\phi=\pi/2$) on an $L_y=4$ cylinder. The pseudo-energies of the ES are sorted according to the $\mathbb{Z}_3$ charge (0 to 2 from top to bottom) and the SU(3) irreps (defined by Dynkin labels) and displayed with decreasing magnitude along the horizontal axis (arbitrary units). The insets show the transfer matrix spectra in the complex plane -- the real (imaginary) part being associated to the correlation length (spatial oscillations of the correlation function). 
We have imposed the three different groups of irreps on the bonds. Only the middle spectrum corresponds to an injective MPS (irreps with $Q=1$), whereas the top and bottom correspond to non-injective MPS obtained by artificially adding a virtual $\mathbf{3}$ or $\overline{\mathbf{3}}$ bond. The exact degeneracies in the top and bottom entanglement spectra (in different $\SU(3)$ sectors) are the signatures of adding this extra virtual irrep. In addition, the occurrence of a transfer matrix eigenvalue in the $(1,1)$ sector on the unit circle points to a non-injective MPS.}
\label{fig:comb_aklt}
\end{figure}

 We have now optimized iMPSs in the three different classes for the $L_y=4$ cylinder. The resulting entanglement spectra and transfer matrix spectra can be found in Fig.~\ref{fig:comb_aklt}. As one can see from these figures, only one of the three choices of virtual irreps gives rise to an injective MPS upon optimization, in this case the $Q=1$ irreps. When choosing the other virtual irreps we find a non-injective MPS, which can be seen from the degeneracies in the entanglement spectrum and the fact that we find a transfer-matrix eigenvalue on the unit circle in a non-trivial sector (depicted as insets in Fig.~\ref{fig:comb_aklt}).

\subsection{Trimerized phase}

We can play the same game in the trimerized phase. In this phase, the ground state has a strong inclination to form singlets on the triangles, and the iMPS geometry will clearly favor the trimerization to happen on the up triangles. The fully-trimerized state (product state of trimers on up-triangles) is represented by the above MPS, with $I_v=I_v'=\mathbf{1}$ on the bonds. Therefore, all iMPSs in this phase are adiabatically connected with this product state and will have virtual irreps with $Q=0$, irrespective of the cylinder circumference.
\par
\begin{figure} \begin{center}
			\includegraphics[width=\columnwidth]{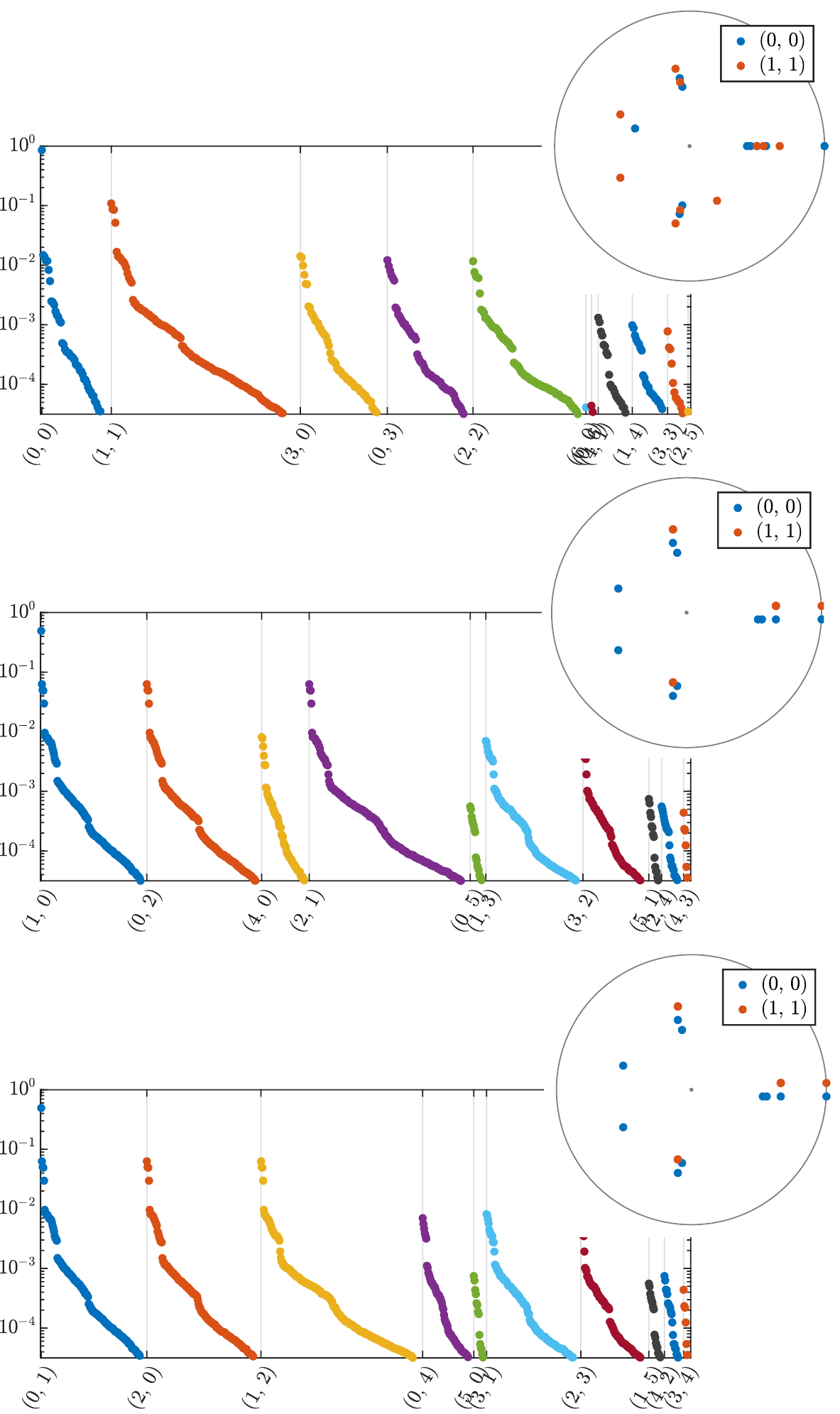}
\end{center} 
\caption{
%\justifying
MPS entanglement (main) and transfer matrix (inset) spectra of an optimized MPS in the trimer phase ($\theta=0$, $\phi=0$) on an $L_y=4$ cylinder (same display as in Fig.~\ref{fig:comb_aklt}), where we have imposed the three different groups of irreps on the bonds. Only the top spectrum corresponds to an injective MPS (irreps with $Q=0$), whereas the lower two panels correspond to non-injective MPS obtained by artificially adding a virtual $\mathbf{3}$ or $\overline{\mathbf{3}}$ bond. Again, the exact degeneracies in these two entanglement spectra are the signatures of adding this extra virtual irrep. The bond dimension of the $Q=0$ MPS is around $\chi\approx7000$.}
\label{fig:comb_trimer}
\end{figure}

 As an illustration, in Fig.~\ref{fig:comb_trimer} we have plotted the MPS entanglement and transfer matrix spectra for the MPS ground state in the point $\theta,\phi=0$ for circumference \mbox{$N=4$}. Clearly, only the choice of $Q=0$ irreps leads to the correct MPS ground state, whereas choosing $Q=\pm1$ leads to non-injective iMPSs.

\subsection{Topological spin liquid phase}
\label{appendix:sub_csl}

\begin{figure} \begin{center}
			\includegraphics[width=\columnwidth]{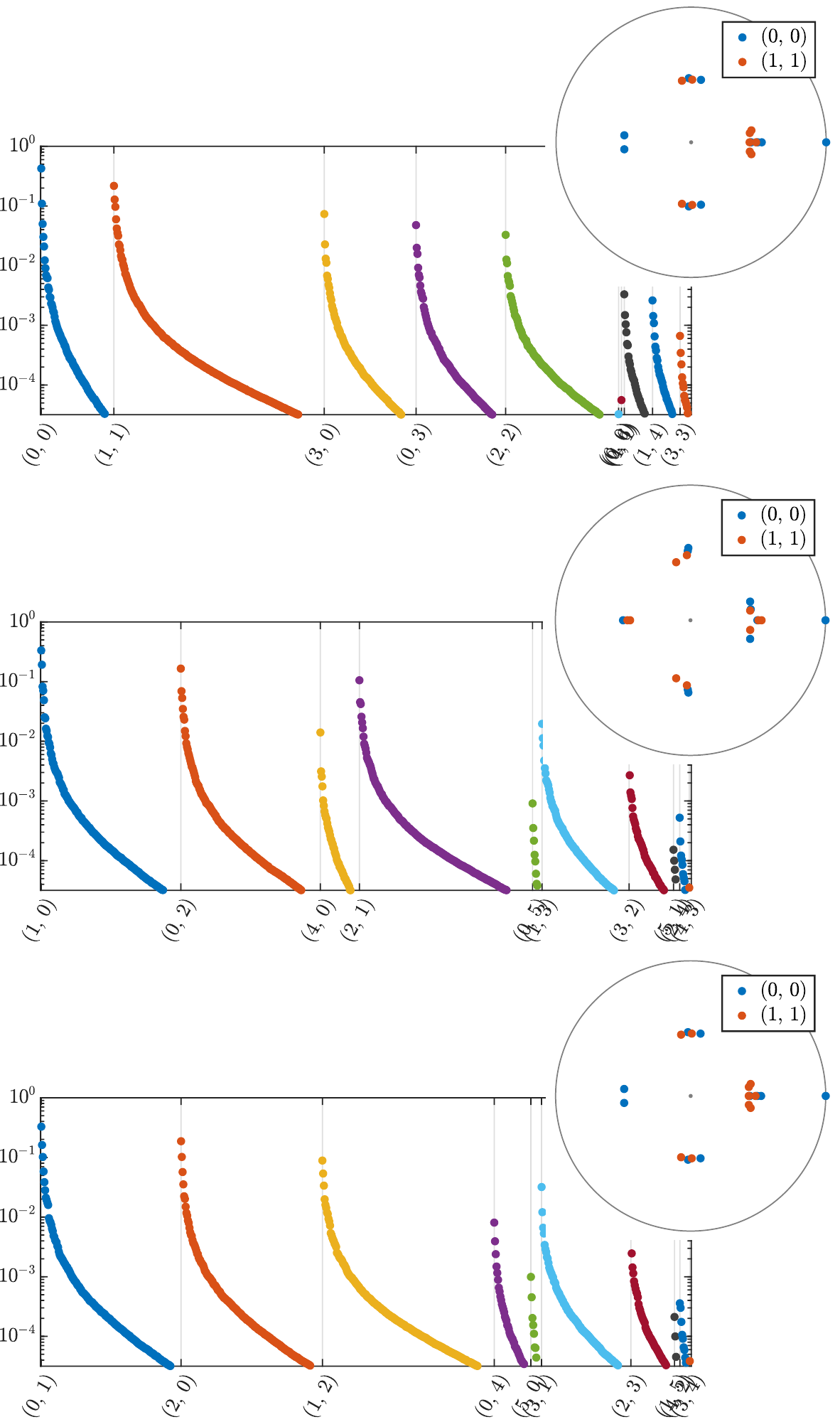}
\end{center} 
\caption{
%\justifying
MPS entanglement (main) and transfer matrix (inset) spectra of an optimized MPS in the TSL phase ($\theta=\frac{\pi}{4}$, $\phi=0$) on an $L_y=4$ cylinder (same display as in Fig.~\ref{fig:comb_aklt}), where we have imposed the three different groups of irreps on the bonds. All three choices give rise to injective MPS (as expected in a TSL phase), and no artificial degeneracies are observed in the entanglement spectra or transfer matrix spectra. The energy densities are almost equal: $e_{Q=0}=-0.7318557278$, $e_{Q=+1}=-0.7317589213$, $e_{Q=-1}=-0.7318473342$ and the total bond dimensions are all around $\chi\approx12000$.}
\label{fig:comb_csl}
\end{figure}

In the spin-liquid phase, the situation is different because we expect three distinct ground states on the infinite cylinder, which are labeled by the $\Z_3$ charge. Indeed, we find that the leading eigenvalues of the MPS entanglement spectrum are the same up to the fourth significant digit in the three charge sectors $Q=0, 1$ and 2 (see Fig.~\ref{fig:comb_csl}). 
The degeneracy is exponential in the circumference of the cylinder.

The 3-fold degenerate nature of the TSL ground state is also corroborated by the leading eigenvalue of the iMPS transfer matrix, which lies on the unit circle in the imaginary plane and is degenerate among all three charge sectors (see insets in Fig.~\ref{fig:comb_csl}).

\subsection{Estimating the gap}
\label{appendix:sub_gap}

In order to estimate the gap, we apply the quasiparticle excitation ansatz. In this context, this boils down to applying as a variational ansatz the state
\begin{equation}
\ket{\Phi_q^s(B)} = \sum_n \e^{iqn} \mydiagram{5},
\end{equation}
which has well-defined momentum $q$ and $\SU(3)$ quantum number $s$. Note that the $\SU(3)$ fusion rules dictate that $s$ should have a quantum number with $Q=0$, i.e. we have $s=\mathbf{1}, \mathbf{8}, \dots$. We can variationally optimize the tensor $B$ for every momentum $q$ and in each sector $s$, yielding a variational dispersion relation. By choosing the shifted boundary conditions, the momentum quantum number follows one continuous line through the 2-D Brillouin zone. Note that, if we have multiple ground states (in the spin liquid phase), we can build domain wall states that carry fractional quantum numbers (i.e., other quantum numbers $s$). The description of these spinon excitations is not further pursued here.
\section{Projected Entangled Simplex States (PESS) and Pair States (PEPS)}
\label{appendix:PESS}

\subsection{General formulation}

\textit{PESS}: The wavefunction for the 1-triangle unit cell is defined as a product of 3 site projectors and 2 trivalent tensors, $(B_a)^{s_a}_{ip}$, $(B_b)^{s_c}_{ql}$, $(B_c)^{s_c}_{rs}$, $(T_d)_{pqr}$, $(T_u)_{sjk}$, given by
\begin{align}
    |\psi(s_a,s_b,s_c)\rangle &= (B_a)^{s_a}_{ip}(T_d)_{pqr}(B_b)^{s_b}_{ql}(B_c)^{s_c}_{rs}(T_u)_{sjk}\notag\\
    &=\vcenter{\hbox{\includegraphics[width=0.35\columnwidth]{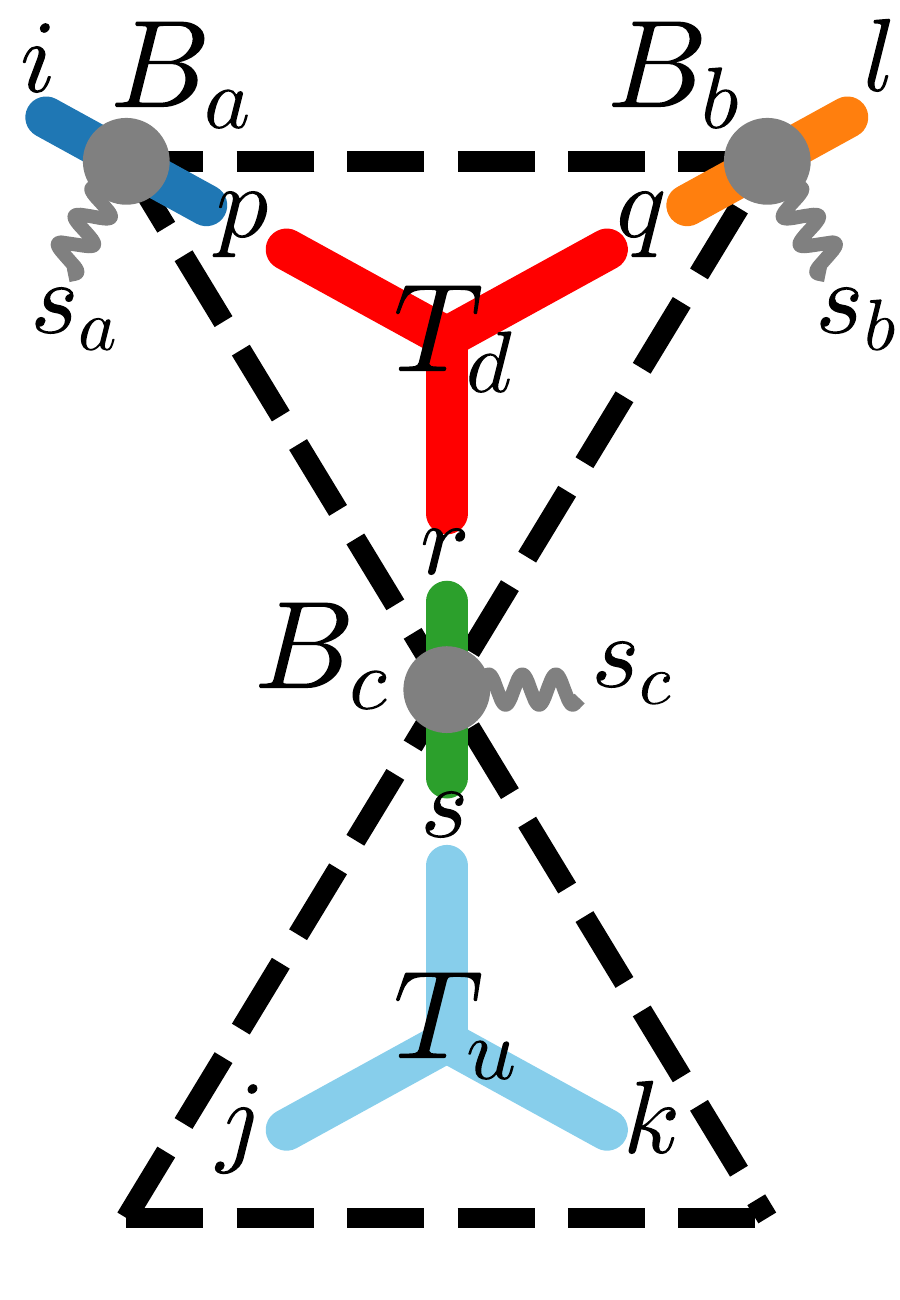}}}.
\end{align}

\textit{PEPS}: In PEPS construction, each down (or up) triangle is considered as the basic building block which tiles the entire lattice. The wavefunction for the three sites in each down (or up) triangle is given by a single rank-5 PEPS tensor which fuses all the physical degrees of freedom together.
\begin{equation}
    |\psi(s_a,s_b,s_c)\rangle = a^{S}_{uldr}, \ S=s_a s_b s_c
\end{equation}

\subsection{1-triangle PEPS phase diagram}
\label{app:1triangle}

\begin{figure}[htb]
	\centering
  		\includegraphics[width=\columnwidth]{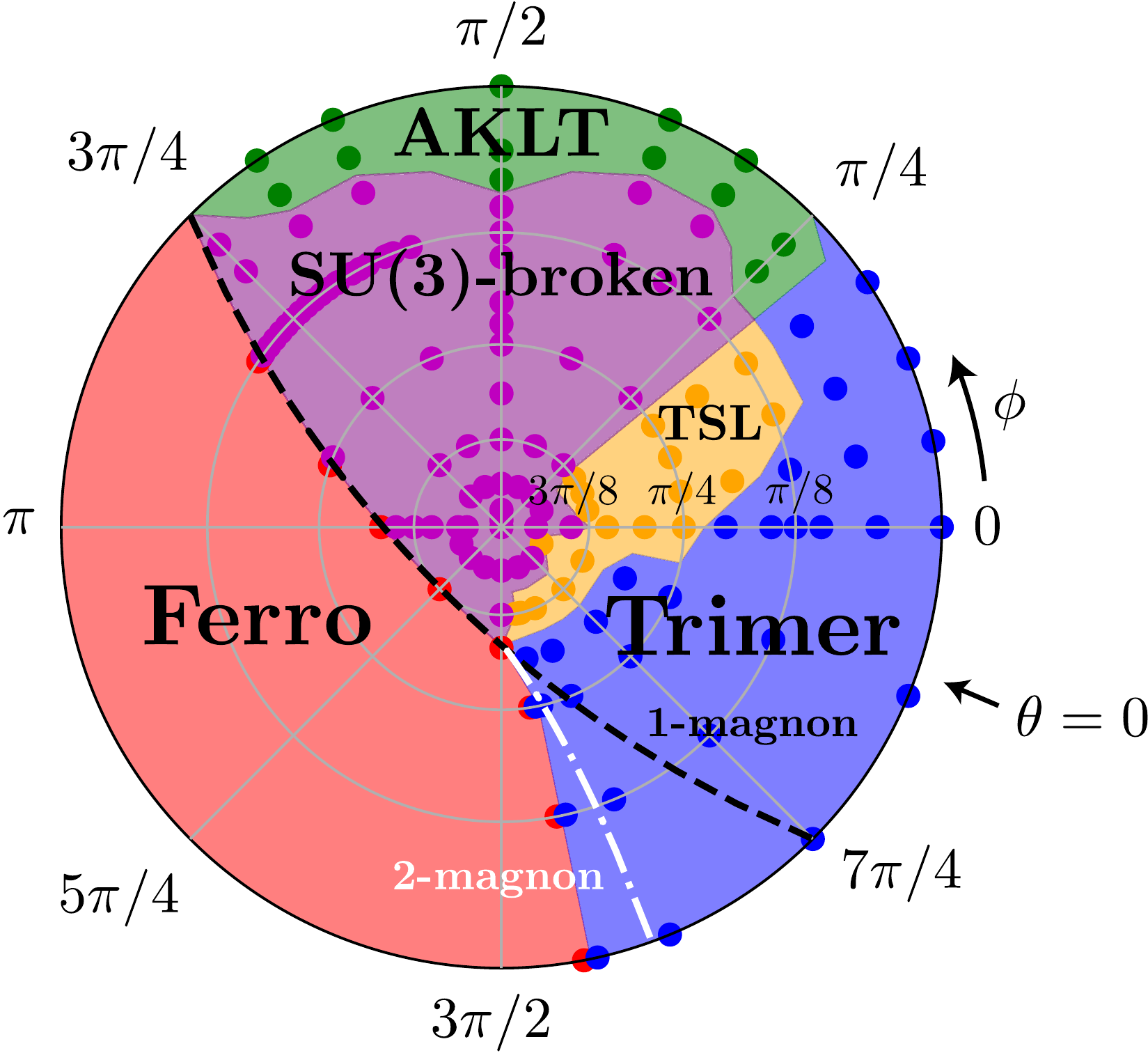}
	\caption{
    %\justifying
    Stereographic projection of the phase diagram (same parametrisation as in Figs.~\ref{fig:phasediag} and \ref{fig:phasediag2}) of the SU(3) chiral antiferromagnet on the kagome lattice obtained by a 1-triangle PEPS ansatz.
		The various phases discussed in the text are represented by dots of different colors. The dashed (dash-dotted) line corresponds to the single (two) magnon instability of the ferromagnetic phase. 
		\label{fig:phasediag3}
}
\end{figure}

In this section, we describe the phase diagram obtained using a 1-triangle iPEPS (see Fig.~\ref{fig:phasediag3}), revealing significant differences compared to the one shown in Fig.~\ref{fig:phasediag2}. 

The criteria for identifying different phases are elaborated as follow. First, states in the SU(3) broken phase (in magenta) has non-zero magnetization (a threshold of $m_{1}=0.01$ is chosen while the maximum allowed value is $m_{0}=2/\sqrt{3}$ for the Ferro states in red). Then, one computes the projection operators onto $\bf1$, $\bf 8$, $\bf8$, $\bf10$ respectively for down and up triangles. If there is a inversion symmetry breaking on up and down triangles, the state is identified as the trimer state. For those states preserving the inversion symmetry, if the two dominant irreducible representations are the two $\bf 8$, the state is identified as the AKLT phase. Otherwise, if the second dominant irreducible representation is $\bf 1$, the state is identified as the CSL state.

\subsection{3-triangle unit cell PESS and PEPS}
\label{app:3triangles}

For the 3-triangle unit cell PESS and PEPS ansatzes, the unit cell is extended along one direction to contain 9 sites (3 triangles). Neither of them, if not explicitly pointed out, impose constraints of point group symmetry on the tensors. For the 3-triangle PESS ansatz, there are three \textbf{independent} sets of PESS tensors for the three triangles, i.e., $5\times 3=15$ PESS tensors -- $\{T_d^\alpha, T_u^\alpha, B_a^\alpha, B_b^\alpha, B_c^\alpha\}$, $\alpha=1,2,3$. For the 3-triangle PEPS ansatz, a product of three \textbf{independent} 1-triangle PEPS tensors, $(a^\alpha)^{S}_{uldr}$, $\alpha=1,2,3$, are used to represent the physical wavefunction for the 9 sites in the 3 triangles.

\begin{figure}[thb]
    \centering
    \includegraphics[width=0.8\columnwidth]{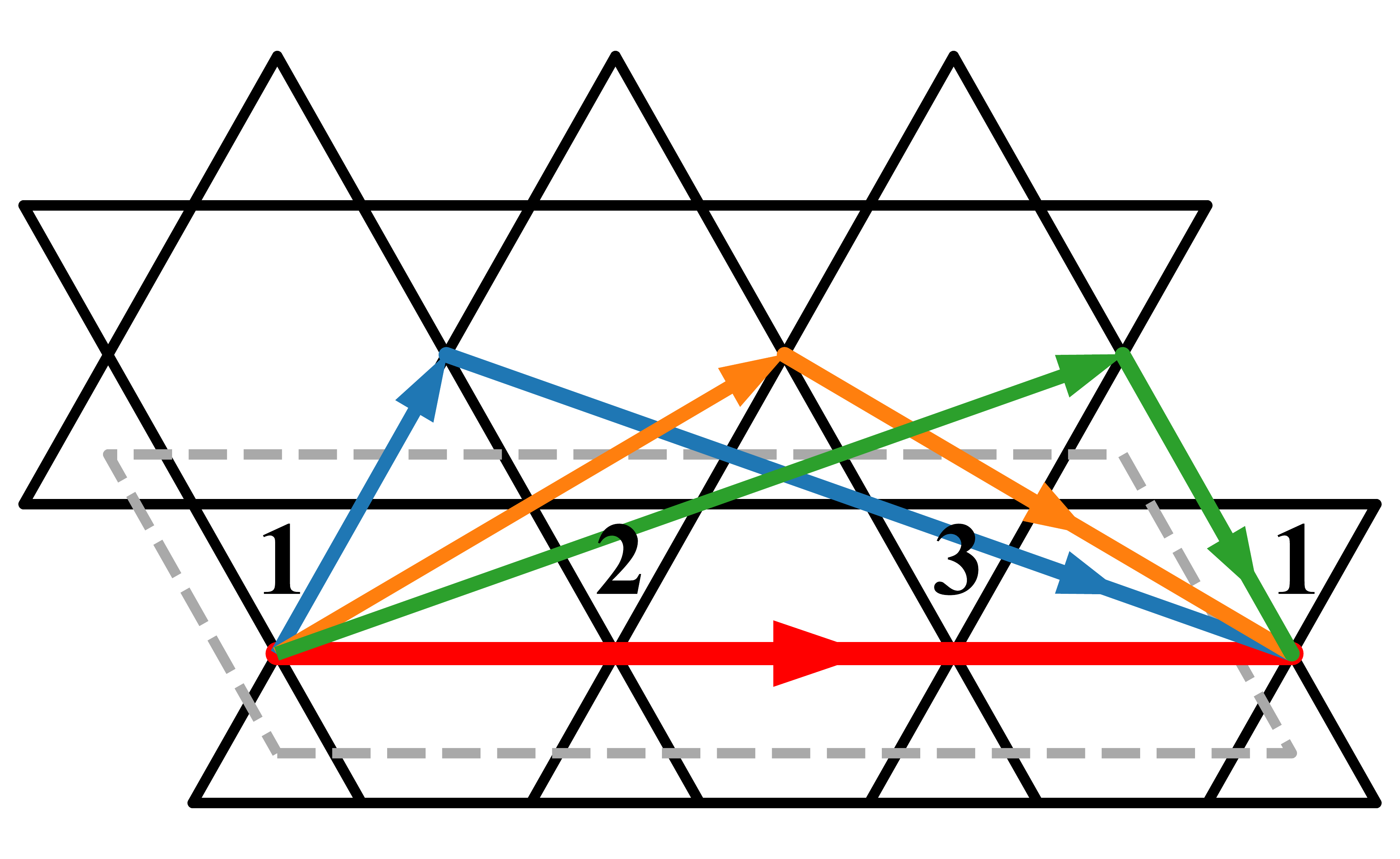}
    \caption{
    %\justifying
    Three choices of lattice vectors that tile the kagome lattice by the 3-triangle unit cell.}
    \label{fig:3-triangle_tiling}
\end{figure}

For the 3-triangle unit cell, there are three choices of the lattice vectors to tile the entire kagome lattice, while two of them are related by mirror symmetry, as shown in Fig.~\ref{fig:3-triangle_tiling}. The tiling with equal length lattice vectors has the C3 rotation symmetry, and thus is denoted by $\sqrt{3}\times\sqrt{3}$ tiling. The other two are simply referred to as $3\times 1$ tiling.

\begin{figure}[thb]
    \centering
    \includegraphics[width=0.4\columnwidth,angle=180]{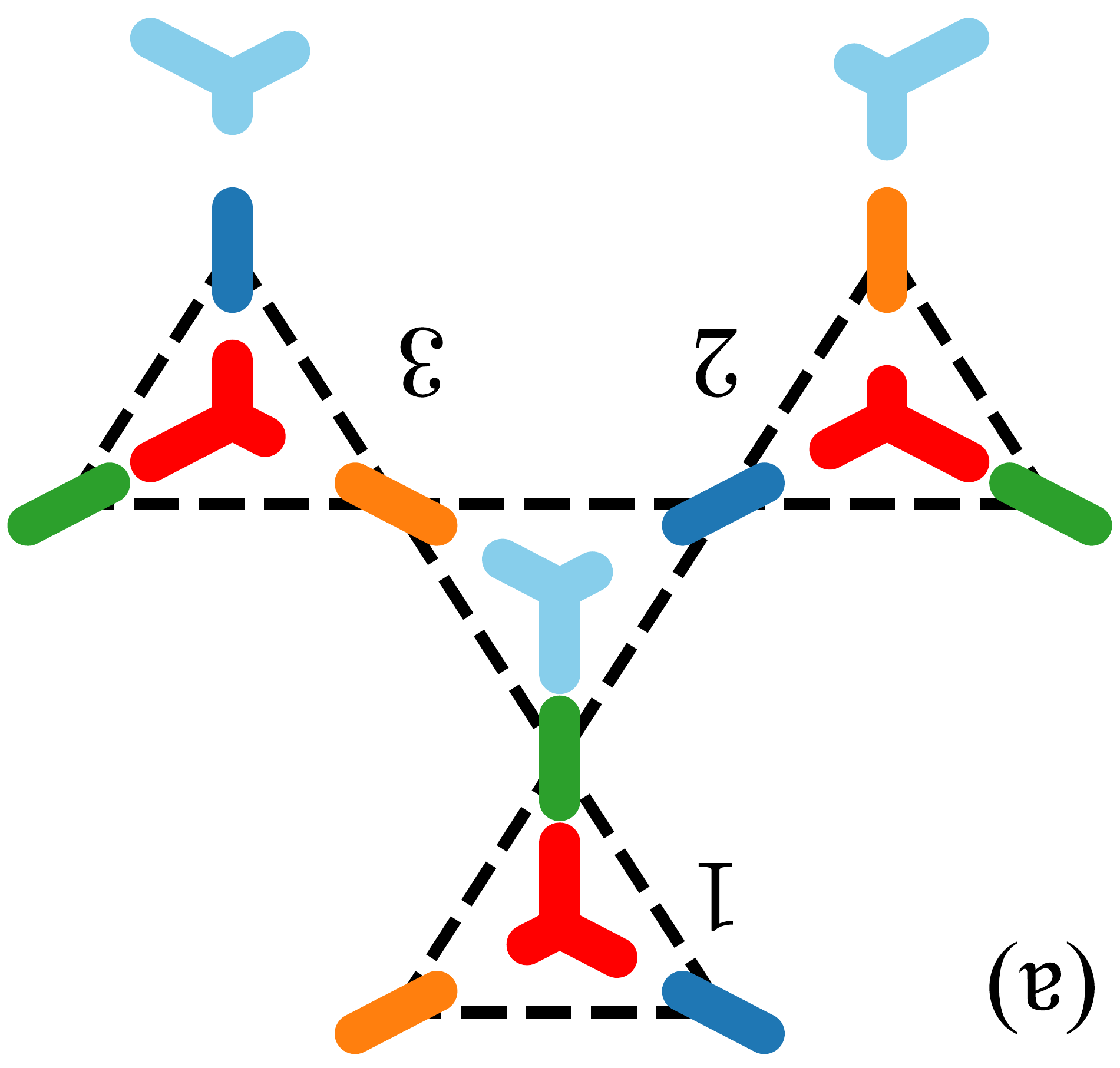}    \hspace{0.1\columnwidth}\includegraphics[width=0.4\columnwidth,angle=180]{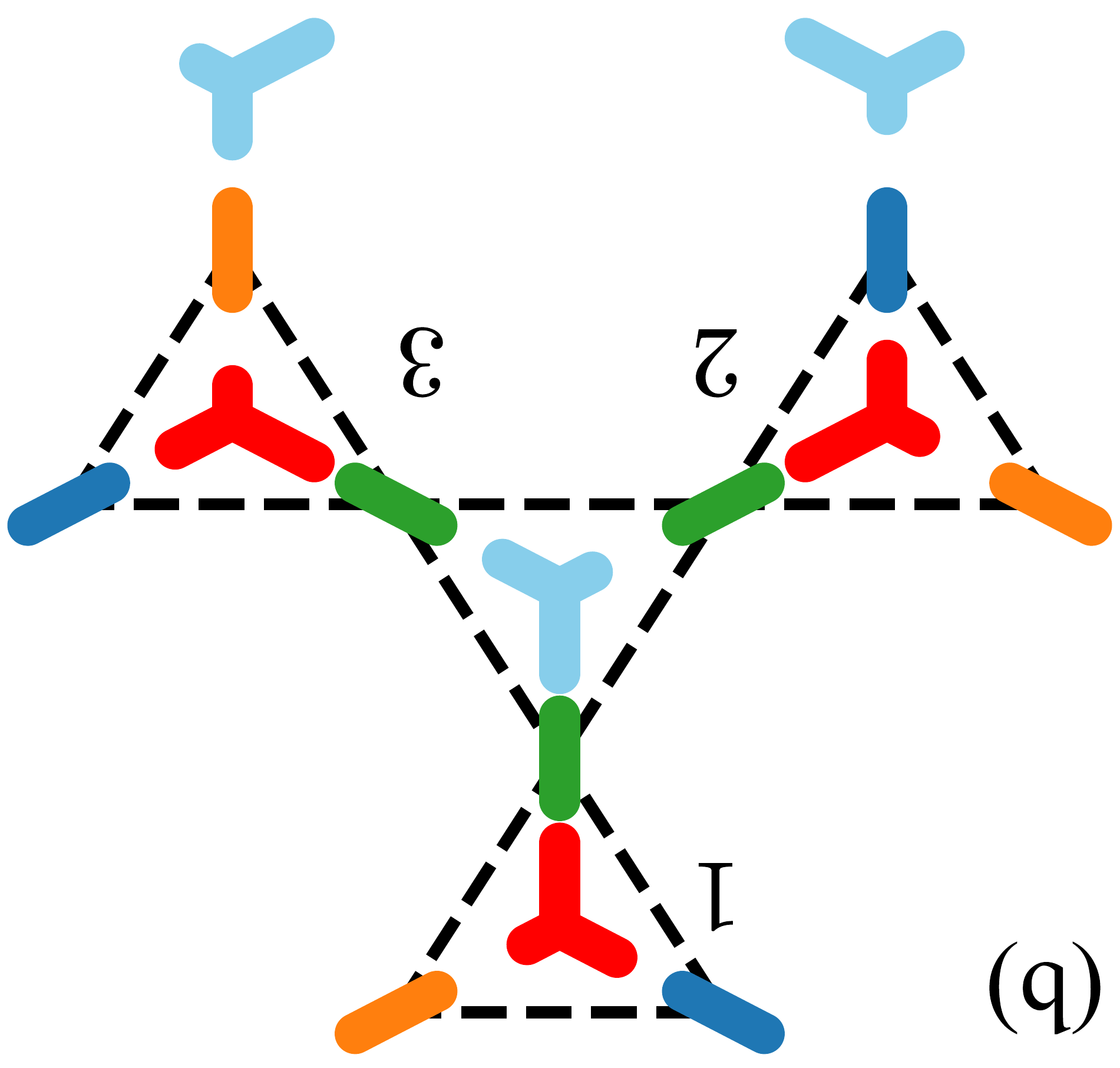}
    \caption{
    %\justifying
    The patterns of the PESS tensors for (a) C3-1 PESS and (b) C3-2 PESS. The bond tensors (site projectors) in the same color are forced to be identical, with the same index always contracted with the same type of trivalent tensor (either $T_d$ or $T_u$). For each trivalent tensor, legs of different lengths stand for different indices. For down or up trivalent tensors in different triangles, the legs of the same length correspond to each other.}
    \label{fig:c3_pess}
\end{figure}

The C3 PESS is a constrained kind of 3-triangle PESS with $\sqrt{3}\times\sqrt{3}$ tiling whose physical wavefunction has C3 lattice rotation symmetry, and is constructed using only one 1-triangle PESS by rotating the PESS tensors for the first triangle to obtain the PESS tensors for the other two. There are two different ways of rotation, which give rise to two different patterns, as shown in Fig.~\ref{fig:c3_pess}. The corresponding relations for PESS tensors in different triangles are given as follow:

\begin{equation}
\text{C3-1 PESS:}\left\{\begin{array}{l}
    (T_d^1)_{pqr}=(T_d^2)_{rpq}=(T_d^3)_{qrp}\\
    (T_u^1)_{sjk}=(T_u^2)_{ksj}=(T_u^3)_{jks}\\
    (B_a^1)_{ip}=(B_b^2)_{lq}=(B_c^2)_{sr}\\
    (B_b^1)_{ql}=(B_c^2)_{rs}=(B_a^3)_{pi}\\
    (B_c^1)_{rs}=(B_a^2)_{pi}=(B_b^3)_{ql}
\end{array}\right.
\end{equation}

\begin{equation}
\text{C3-2 PESS:}\left\{\begin{array}{l}
    (T_d^1)_{pqr}=(T_d^2)_{qrp}=(T_d^3)_{rpq}\\
    (T_u^1)_{sjk}=(T_u^2)_{sjk}=(T_u^3)_{sjk}\\
    (B_a^1)_{ip}=(B_c^2)_{sr}=(B_b^2)_{lq}\\
    (B_b^1)_{ql}=(B_a^2)_{pi}=(B_c^3)_{rs}\\
    (B_c^1)_{rs}=(B_b^2)_{ql}=(B_a^3)_{pi}
    \end{array}\right.
\end{equation}

% C3-1
% \begin{align}
%     (T_d^1)_{pqr}=(T_d^2)_{rpq}=(T_d^3)_{qrp}\notag\\
%     (T_u^1)_{sjk}=(T_u^2)_{ksj}=(T_u^3)_{jks}\notag\\
%     (B_a^1)_{ip}=(B_b^2)_{lq}=(B_c^2)_{sr}\notag\\
%     (B_b^1)_{ql}=(B_c^2)_{rs}=(B_a^3)_{pi}\notag\\
%     (B_c^1)_{rs}=(B_a^2)_{pi}=(B_b^3)_{ql}
% \end{align}
% C3-2
% \begin{align}{l}
%     (T_d^1)_{pqr}=(T_d^2)_{qrp}=(T_d^3)_{rpq}\notag\\
%     (T_u^1)_{sjk}=(T_u^2)_{sjk}=(T_u^3)_{sjk}\notag\\
%     (B_a^1)_{ip}=(B_c^2)_{sr}=(B_b^2)_{lq}\notag\\
%     (B_b^1)_{ql}=(B_a^2)_{pi}=(B_c^3)_{rs}\notag\\
%     (B_c^1)_{rs}=(B_b^2)_{ql}=(B_a^3)_{pi}
% \end{align}

\subsection{Lattice symmetry breaking in SU(3)-broken phase?}

\begin{figure}[thb]
    \centering
    \includegraphics[width=\columnwidth]{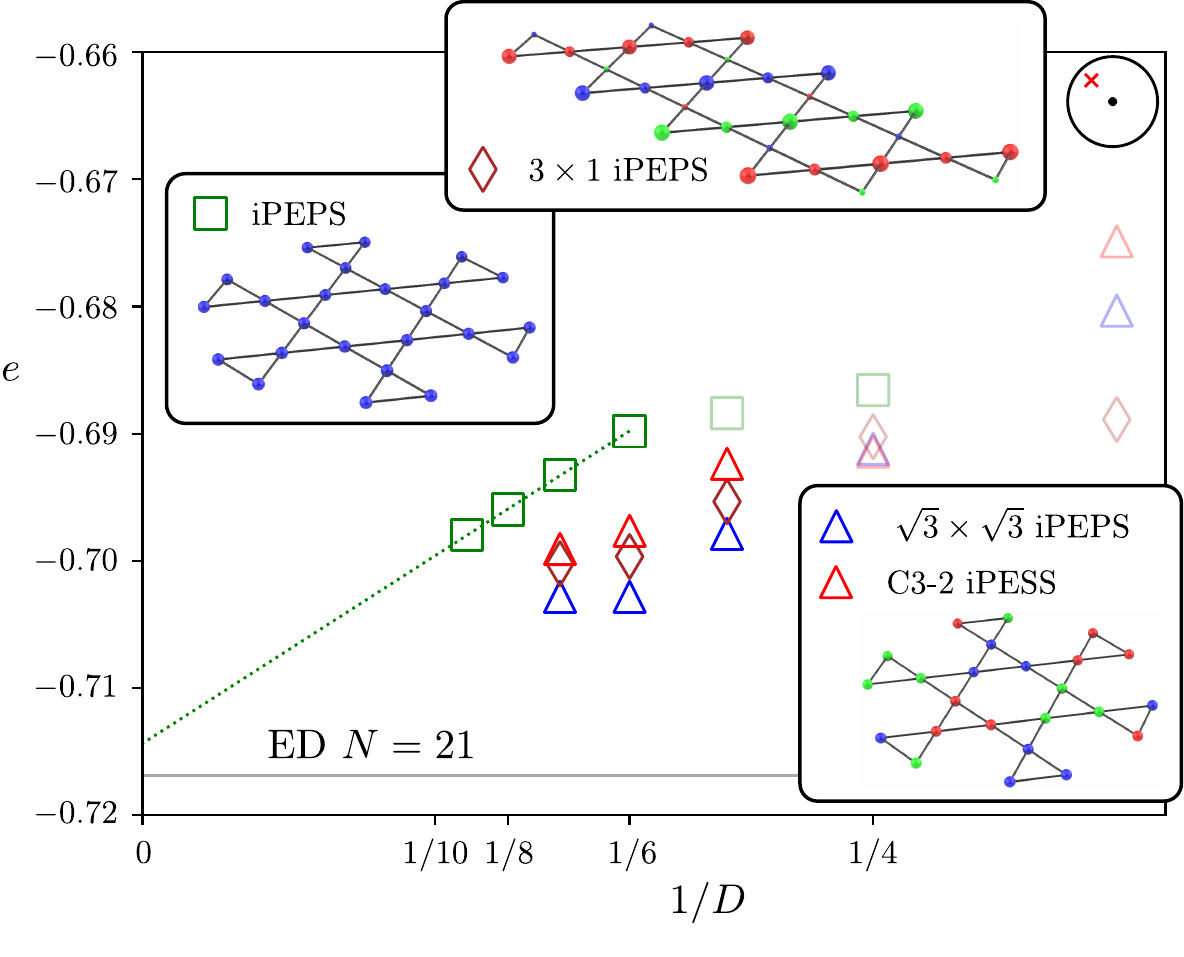}
    \caption{
    %\justifying
    Scaling of energetics for iPEPS and iPESS wave functions with respect to bond dimensions at $(\phi,\theta)=(\frac{3\pi}{4}, \frac{\pi}{8})$. The corresponding spatial patterns of different ansatzes only appear with large bond dimensions (indicated by opaque markers). 1-triangle iPEPS wave function (green squares) gives a uniform pattern with all the spins pointing towards the same direction. $\sqrt{3}\times\sqrt{3}$ iPEPS and C3-2 iPESS (blue and red triangles) wave functions both give a C3-rotation symmetry broken (with respect to the center of each hexagon) pattern. $3\times1$ iPEPS wave function (brown diamonds) gives a partially lattice translation symmetry broken pattern along the direction in which the unit cell is enlarged.}
    \label{fig:e_scaling_phi=0.75pi_theta=0.125pi}
\end{figure}

In Fig.~\ref{fig:cut0125theta} of the main text, we have shown the C3-2 iPESS results (red triangles), while having fewer variational parameters than the 1-triangle iPEPS ansatz with the same bond dimension, can establish lower energy states with additional lattice symmetry breaking patterns. Here, we make a more detailed scaling analysis of the energetics at one point, $(\phi,\theta)=(\frac{3\pi}{4},\frac{\pi}{8})$, where the potential lattice symmetry breaking happens, as shown by Fig.~\ref{fig:e_scaling_phi=0.75pi_theta=0.125pi}. First, one can see that the extrapolation of the energies from 1-triangle iPEPS wave function already gives a value which is very close to the $N=21$ ED result (with a difference smaller than $3\times10^{-3}$). Second, as the bond dimension increases, one can see the energy gap between the uniform states and lattice symmetry broken states decreases. Based on these facts, we tend to attribute the lattice symmetry breaking we observe to the finite bond dimension effects. In other words, with small bond dimensions (low entanglement), the states gain more energy by breaking the lattice symmetry. A similar phenomenon has also been observed in an SU(3) model on the honeycomb lattice~\cite{Corboz2013}. But still, to clear up the issue, one needs to go for larger bond dimensions, which unfortunately goes beyond our current computational capability.

\subsection{SU(3)-symmetric PESS}
The SU(3) symmetric PESS is a constrained family of 1-site PESS, where each tensor is invariant under SU$(3)$ symmetry. In addition, for the AKLT phase and the CSL phase, the SU(3) PESS is further constrained by the lattice point group symmetry.

In practice, first the relevant SU(3) irreps in the virtual spin are found using the simple update method~\cite{Jiang2008}. Then a tensor classification is carried out for both the site projectors and trivalent tensors. The classification scheme follows from Ref.~\cite{Jiang2015, Mambrini2016}, which was recently adapted to the kagome lattice~\cite{Niu2022}.

\begin{table*}[htb]
	\centering
	\begin{tabular}{|c|l|}
		\hline \hline
		$L_0$  & Irreps / Multiplicities	\\	\hline
		$0$	&  \Yvcentermath1$\su{1}{0}{1}$				\\ \hline				
		$1$ &  \Yvcentermath1$\su{1}{2,1}{8}$ 		\\ \hline	
		$2$	&  \Yvcentermath1$\su{1}{0}{1}\oplus\su{2}{2,1}{8}$ \\ \hline	
		$3$	&  \Yvcentermath1$\su{2}{0}{1}\oplus\su{3}{2,1}{8}\oplus\su{1}{3}{10}\oplus\su{1}{3,3}{\overline{10}}$ 								\\ \hline	
		$4$	&  \Yvcentermath1$\su{3}{0}{1}\oplus\su{6}{2,1}{8}\oplus\su{1}{3}{10}\oplus\su{1}{3,3}{\overline{10}}\oplus\su{1}{4,2}{27}$ \\ \hline
		$5$	&  \Yvcentermath1$\su{4}{0}{1}\oplus\su{10}{2,1}{8}\oplus\su{3}{3}{10}\oplus\su{3}{3,3}{\overline{10}}\oplus\su{2}{4,2}{27}$ \\ \hline
		$6$	&  \Yvcentermath1$\su{8}{0}{1}\oplus\su{16}{2,1}{8}\oplus\su{5}{3}{10}\oplus\su{5}{3,3}{\overline{10}}\oplus\su{5}{4,2}{27}$ \\ \hline
		$7$	&  \Yvcentermath1$\su{10}{0}{1}\oplus\su{27}{2,1}{8}\oplus\su{9}{3}{10}\oplus\su{9}{3,3}{\overline{10}}\oplus\su{8}{4,2}{27}\oplus\su{1}{5,4}{\overline{35}}\oplus\su{1}{5,1}{35}$ \\ \hline
		\hline
	\end{tabular}
 \caption{\label{tab:wzw-Q0}
 %\justifying
		Conformal tower in the $Q=0$ topological sector, originating from $\protect\su{0}{0}{1}$ -- reproduced for convenience from Table VI of Ref.~\cite{Chen2021}. For the other SU(3)$_1$ conformal towers in the $Q=1$ and $Q=2$ topological sectors see Table VII of Ref.~\cite{Chen2021}.
  }
\end{table*}

\begin{table*}[htb]
    \centering
    \begin{tabular}{|c|l|}
		\hline \hline
		$L_0$  & Irreps / Multiplicities	\\	\hline
		$0$	& \Yvcentermath1$\su{1}{1}{3}\oplus\su{1}{2,2}{\bar{6}}$				\\ \hline				
		$1$ & \Yvcentermath1$\su{2}{1}{3}\oplus\su{1}{2,2}{\bar{6}}\oplus\su{1}{3,1}{15}$ 		\\ \hline	
		$2$	& \Yvcentermath1$\su{3}{1}{3}\oplus\su{3}{2,2}{\bar{6}}\oplus\su{2}{3,1}{15}\oplus\su{1}{4,3}{24}$ \\ \hline	
		$3$	& \Yvcentermath1$\su{6}{1}{3}\oplus\su{5}{2,2}{\bar{6}}\oplus\su{5}{3,1}{15}\oplus\su{2}{4,3}{24}$ 								\\ \hline	
		$4$	& \Yvcentermath1$\su{10}{1}{3}\oplus\su{10}{2,2}{\bar{6}}\oplus\su{1}{4}{15'}\oplus\su{9}{3,1}{15}\oplus\su{4}{4,3}{24}\oplus\su{1}{5,2}{42}$ \\ \hline
		$5$	& \Yvcentermath1$\su{17}{1}{3}\oplus\su{16}{2,2}{\bar{6}}\oplus\su{2}{4}{15'}\oplus\su{17}{3,1}{15}\oplus\su{1}{5,5}{21}\oplus\su{8}{4,3}{24}\oplus\su{2}{5,2}{42}$ \\ \hline
		$6$	& \Yvcentermath1$\su{27}{1}{3}\oplus\su{28}{2,2}{\bar{6}}\oplus\su{4}{4}{15'}\oplus\su{29}{3,1}{15}\oplus\su{1}{5,5}{21}\oplus\su{15}{4,3}{24}\oplus\su{5}{5,2}{42}\oplus\su{1}{6,4}{60}$ \\ \hline
		$7$	& \Yvcentermath1$\su{43}{1}{3}\oplus\su{43}{2,2}{\bar{6}}\oplus\su{8}{4}{15'}\oplus\su{50}{3,1}{15}\oplus\su{3}{5,5}{21}\oplus\su{26}{4,3}{24}\oplus\su{10}{5,2}{42}\oplus\su{2}{6,4}{60}$ \\ \hline
		\hline
	\end{tabular}
   \caption{
   %\justifying
   Conformal tower in the $Q=1$ topological sector: tower originating from $\protect\su{0}{1,1}{\bar{3}}$ [$\otimes\protect\su{0}{1,1}{\bar{3}}$ ].}
    \label{tab:wzw-Q1}
\end{table*}

\begin{table*}[]
    \centering
    \begin{tabular}{|c|l|}
		\hline \hline
		$L_0$  & Irreps / Multiplicities	\\	\hline
		$0$	&  \Yvcentermath1$\su{1}{1,1}{\bar{3}}\oplus\su{1}{2}{6}$				\\ \hline				
		$1$ &  \Yvcentermath1$\su{2}{1,1}{\bar{3}}\oplus\su{1}{2}{6}\oplus\su{1}{3,2}{\overline{15}}$ 		\\ \hline	
		$2$	&  \Yvcentermath1$\su{3}{1,1}{\bar{3}}\oplus\su{3}{2}{6}\oplus\su{2}{3,2}{\overline{15}}\oplus\su{1}{4,1}{\overline{24}}$ \\ \hline	
		$3$	&  \Yvcentermath1$\su{6}{1,1}{\bar{3}}\oplus\su{5}{2}{6}\oplus\su{5}{3,2}{\overline{15}}\oplus\su{2}{4,1}{\overline{24}}$ 								\\ \hline	
		$4$	&  \Yvcentermath1$\su{10}{1,1}{\bar{3}}\oplus\su{10}{2}{6}\oplus\su{9}{3,2}{\overline{15}}\oplus\su{1}{4,4}{\overline{15}'}\oplus\su{4}{4,1}{\overline{24}}\oplus\su{1}{5,3}{\overline{42}}$ \\ \hline
		$5$	&  \Yvcentermath1$\su{17}{1,1}{\bar{3}}\oplus\su{16}{2}{6}\oplus\su{17}{3,2}{\overline{15}}\oplus\su{2}{4,4}{\overline{15}'}\oplus\su{1}{5}{\overline{21}}\oplus\su{8}{4,1}{\overline{24}}\oplus\su{2}{5,3}{\overline{42}}$ \\ \hline
		$6$	&  \Yvcentermath1$\su{27}{1,1}{\bar{3}}\oplus\su{28}{2}{6}\oplus\su{29}{3,2}{\overline{15}}\oplus\su{4}{4,4}{\overline{15}'}\oplus\su{1}{5}{\overline{21}}\oplus\su{15}{4,1}{\overline{24}}\oplus\su{5}{5,3}{\overline{42}}\oplus\su{1}{6,2}{\overline{60}}$ \\ \hline
		$7$	&  \Yvcentermath1$\su{43}{1,1}{\bar{3}}\oplus\su{43}{2}{6}\oplus\su{50}{3,2}{\overline{15}}\oplus\su{8}{4,4}{\overline{15}'}\oplus\su{3}{5}{\overline{21}}\oplus\su{26}{4,1}{\overline{24}}\oplus\su{10}{5,3}{\overline{42}}\oplus\su{2}{6,2}{\overline{60}}$ \\ \hline
		\hline
	\end{tabular}
   \caption{
   %\justifying
   Conformal tower in the $Q=2$ topological sector: tower originating from $\protect\su{0}{1}{3}$ [$\otimes\protect\su{0}{1}{3}$ ]. }
    \label{tab:wzw-Q2}
\end{table*}

\clearpage
\bibliography{bibliography}
\end{document}